\documentclass[pra, a4paper, preprint, showpacs,superscriptaddress]{revtex4-1}
\usepackage{amsmath,amscd,amsfonts,amssymb, amsbsy, color,bbm, bm}
\usepackage{graphicx}
\usepackage{longtable}
\begin{document}
\title{From quantum stochastic differential equations to Gisin-Percival state diffusion}  
\author{K. R. Parthasarathy}\email{krp@isid.ac.in}
\affiliation{Indian Statistical Institute, Theoretical Statistics and Mathematics Unit,Delhi Centre,
7 S.~J.~S. Sansanwal Marg, New Delhi 110 016, India} 
\author{A. R. Usha Devi}\email{arutth@rediffmail.com} 
\affiliation{Department of Physics, Bangalore University, 
Bangalore-560 056, India}
\affiliation{Inspire Institute Inc., Alexandria, Virginia, 22303, USA.}
\date{\today}
\begin{abstract}
Starting from the quantum stochastic differential equations of Hudson and Parthasarathy (Comm. Math. Phys. {\bf 93}, 301 (1984)) and exploiting the Wiener-It{\^o}-Segal isomorphism between the Boson Fock reservoir space $\Gamma(L^2(\mathbb{R}_+)\otimes (\mathbb{C}^{n}\oplus \mathbb{C}^{n}))$ and the Hilbert space $L^2(\mu)$, where $\mu$ is the Wiener  probability measure of a complex $n$-dimensional 
vector-valued standard Brownian motion $\{\mathbf{B}(t), t\geq 0\}$, we  derive a non-linear stochastic Schr{\"o}dinger equation describing a classical diffusion of states of a quantum system,  driven by the Brownian motion $\mathbf{B}$. Changing this Brownian motion by an appropriate Girsanov transformation, we arrive at the Gisin-Percival state diffusion equation (J. Phys. A {\bf 167}, 315 (1992)). This  approach also yields an explicit solution of the Gisin-Percival equation, in terms of the Hudson-Parthasarathy unitary process and a radomized Weyl displacement process. Irreversible dynamics of system density operators described by the well-known Gorini-Kossakowski-Sudarshan-Lindblad master equation is unraveled by coarse-graining over the Gisin-Percival quantum state trajectories. 
\end{abstract}
\maketitle 
\section{Introduction} 

Irreversible dynamics of states and observables of a quantum system $S$ is usually described by a one parameter semigroup 
$\{T_t,t\geq 0\}$ of unital completely positive maps on the algebra $\mathcal{B}(\mathcal{H}_S)$ of all bounded operators on the associated system Hilbert space $\mathcal{H}_S$. Such a semigroup is called a {\it quantum dynamical semigroup}. When this semigroup is uniformly continuous, its infinitesimal generator was completely described by Gorini, Kossakowski and Sudarshan~\cite{Gorini} when $\mathcal{H}_S$ is finite dimensional and by Lindblad~\cite{Lindblad} in the general case. We call it the GKSL generator, usually denoted by $\mathcal{L}$. The form of this generator $\mathcal{L}$ becomes meaningful even when the operators entering the description of $\mathcal{L}$ may be unbounded~\cite{Fan, Fan&Wills, Fan13, KRPG} and can give rise to dynamical semigroups, which are not necessarily uniformly continuous. Since the discovery of the form of the generator $\mathcal{L}$, there have been attempts to understand the stochastic processes from which $\mathcal{L}$ arises. This has mainly given rise to two different approaches of constructing processes leading to the generator $\mathcal{L}$. 

Starting with the 1984 paper~\cite{HP} of Hudson and Parthasarathy (HP), there has evolved a Boson Fock space stochastic calculus for operator-valued processes in $\mathcal{H}_S\otimes\mathcal{H}_R$, where $\mathcal{H}_R$ is an appropriate Boson Fock space associated with a reservoir $R$ (also called {\it bath} or {\it noise}). Such a stochastic calculus is equipped with a {\it quantum It{\^o} formula}~\cite{HP, KRP} leading to a theory of {\it quantum stochastic differential equations}. This enables, in particular, the construction of {\it unitary operator-valued processes} $\{U(t), t\geq 0\}$ satisfying a quantum stochastic differential equation in $\mathcal{H}_S\otimes\mathcal{H}_R$. It turns out that for a given GKSL generator $\mathcal{L}$, there exists a canonical unitary operator-valued process in $\mathcal{H}_S\otimes\mathcal{H}_R$ obeying a quantum stochastic differential equation of the exponential type and satisfying the identity 
$$\langle \phi\vert T_t(X)\vert \chi \rangle=\langle \phi\otimes \Omega_0\vert U(t)^\dag\, (X\otimes I_R)\, U(t)\,\vert \chi\otimes \Omega_0 \rangle$$ 
for all $X\in \mathcal{B}(\mathcal{H}_S)$, $\phi,\, \chi\in \mathcal{H}_S$, where  $I_R$ is the identity operator in $\mathcal{H}_R$, $\Omega_0$ denotes the  Boson Fock vacuum state,  and $\{T_t=e^{t\,\mathcal{L}}, t\geq 0\}$, the dynamical semigroup with generator $\mathcal{L}$. In other words,  $\{T_t,t\geq 0\}$ has been dilated to a Heisenberg evolution by the unitary operator-valued process $\{U(t), t\geq 0\}$. 

On the other hand, in their 1992 paper~\cite{Gisin} Gisin and Percival explore the possibility of constructing the dynamical semigroup $\{T_t,t\geq 0\}$ with GKSL generator $\mathcal{L}$ through classical diffusion processes, with  values on the unit sphere of the system Hilbert space $\mathcal{H}_S$, driven by a complex vector-valued standard Brownian motion $\{\mathbf{B}(t), t\geq 0\}$, with its Wiener probability measure $\mu$, on the space of paths. They arrive at a non-linear diffusion equation on the unit sphere involving the differentials $d\mathbf{B}(t)$ and $dt$, with diffusion and drift coefficients depending on the operator parameters 
describing $\mathcal{L}$. Such classical stochastic differential equations for processes with values in the unit sphere of $\mathcal{H}_S$
 are called {\it stochastic Schr{\"o}dinger equations}.  For any initial state $\vert \phi_0\rangle$ in $\mathcal{H}_S$, the Gisin-Percival stochastic Schr{\"o}dinger equation determines a {\it trajectory}   
$\{\vert\Psi_t(\mathbf{B})\rangle,\ t\geq 0\}$ of pure states in $L^2(\mu)\otimes \mathcal{H}_S$, which is driven by complex vector-valued Brownian noise $\mathbf{B}$. 
The system density operator $\rho_t$, obtained after coarse graining over these diffusive trajectories~\cite{Bar1, Bar2},
$$\rho_t=\int\, \vert \Psi_t(\mathbf{B})\rangle\langle 
\Psi_t(\mathbf{B})\vert\, \mu(d\mathbf{B})$$
 obeys a GKSL master equation~\cite{Gorini, Lindblad}. This determines the irreversible dynamics of states and observables 
in $\mathcal{H}_S$. In other words, pure state solutions of stochastic Schr{\"o}dinger equations can be  employed   effectively in studying open system dynamics.  Non-linear stochastic Schr{\"o}dinger equations have gained importance from various physical and mathematical  
perspectives~\cite{Gisin,Bar1,Bar2,Ghir1,Dio1,Bel1,Ghir2,Gaterek,Bel2,Hol,Adl1,GisinRigo,Dio2,Adl2,Gough,Massen,Bassi}. They were initially proposed~\cite{Dio1} as  stochastic non-linear modifications of the Schr{\"o}dinger equation, as an attempt to address the quantum measurement problem~\cite{Ghir1,Dio1,Ghir2,Adl1,Dio2,Adl2,Bassi}. It has also been recognized that the use of pure states, instead of density matrices, is advantageous in speeding up  computer simulations~\cite{Carm,Molmer1,Molmer2}.      

The main goal of this paper is to construct the Gisin-Percival diffusion of states from the quantum stochastic differential equation of HP~\cite{HP, KRP}, by exploiting the Wiener-It{\^o}-Segal isomorphism~\cite{Wiener,Ito,Segal} between the reservoir Boson Fock space $\mathcal{H}_R$ and the Hilbert space $L^2(\mu)$, with $\mu$ being the Wiener probability measure on the space of paths of a vector-valued Brownian motion. One of the striking features of our derivation is an explicit and simple realization of a solution of the Gisin-Percival equation in terms of an HP unitary process and a randomized Weyl displacement process. Randomized Weyl displacement operators introduced here are themselves unitary  and they are stochastic generalizations of the well known Weyl displacement operators of classical quantum theory. 
 
Our paper is organized in the form of seven sections. Section II contains a discussion on discrete time irreversible dynamics of a finite $d$-level quantum system $S$.  This is intended to prepare a necessary groundwork for its natural adaptation to  continuous time noisy evolution, as formulated by HP~\cite{HP, KRP}. Section III presents a brief account of HP quantum stochastic calculus. A description of  {\it noisy Schr{\"o}dinger unitary evolutions} in terms of quantum stochastic differential equations is presented here. We describe,  how a   unitary operator-valued process $\{U(t), t\geq 0\}$ obeying a quantum stochastic differerntial equation  in $\mathcal{H}_S\otimes \mathcal{H}_R$, leads to  the quantum dynamical semigroup $\{T_t, t\geq 0\}$, with GKSL generator $\mathcal{L}$.  Invariance  properties of the GKSL generator $\mathcal{L}$ under  unitary Weyl displacement process and second quantized  unitary operator-valued process is discussed  in Section IV. The basic notions of the Wiener-It{\^o}-Segal isomorphism between the reservoir space $\mathcal{H}_R$ and the Hilbert space $L^2(\mu)$ of norm square integrable functions with respect to the Wiener probability measure $\mu$ of a vector-valued Brownian motion are presented in Section V. Starting from an HP quantum stochastic differential equation,  Gisin-Percival~\cite{Gisin} quantum state diffusion equation is derived in Section VI. A brief summary of our results is given in Section VII.       
 
\section{The case of irreversible discrete time dynamics of finite $d$-level systems}

Consider a finite $d$-level system $S$ in a Hilbert space $\mathcal{H}_S$. Let $T$ be a unital completely positive map on the algebra  
$\mathcal{B}(\mathcal{H}_S)$ of all bounded operators in $\mathcal{H}_S$. Then the sequence $\{T^0, T^1, T^2, T^3,\cdots\}$ determines a quantum dynamical semigroup. Thanks to the Stinespring's theorem, one can construct a finite probability space 
$(\mathbb{X},\nu)$ with $\mathbb{X}=\{0,1,2,\cdots k-1\}$,  $\nu$ being the uniform distribution with mass $1/k$ at each $x\in\mathbb{X}$, and an orthonormal basis $\{\vert x\rangle, x\in \mathbb{X}\}$ in the Hilbert space $L^2(\nu)$, such that  $\vert 0\rangle$ is the constant function with value unity at every $x$ in $\mathbb{X}$ and a unitary operator $U$ in $\mathcal{H}_S\otimes L^2(\nu)$ determined by 
\begin{equation}
U\, \vert \phi  \otimes  x\rangle = 
\sum_{y\in \mathbb{X}}\, \left( L_{yx}\,\vert \phi\rangle\right)\otimes \vert y\rangle,\ \  
\forall\,\, \vert\phi\rangle\,\in \mathcal{H}_S,\   x\, \in\mathbb{X}
\end{equation} 
with $L_{yx}$ being operators in $\mathcal{H}_S$ for all  $x,y\in \mathbb{X}$, so that 
\begin{equation}
T(X)=\sum_{y\in \mathbb{X}} L_{y0}^\dag\, X\, L_{y0},\ \  \forall\ X\in \mathcal{B}(\mathcal{H}_S).  
\end{equation} 
 In particular, $\sum_y\, L^\dag_{y0}\, L_{y0}=I_S$, where $I_S$ is the identity operator in $\mathcal{H}_S$. 
Denoting $\mathbb{N}=\{ 1,2,\cdots\}$ and the countable product probability space 
$$(\Omega, \mu)=(\mathbb{X}, \nu)^{\otimes \mathbb{N}},$$
where any sample point $\omega\in \Omega$ is a discrete trajectory 
\begin{equation}
\omega=\{x_1,x_2,\cdots, x_n,\cdots\},\  
\end{equation} 
with  $x_1,x_2,\cdots\in\mathbb{X}$ being independently and identically distributed with uniform distribution $\nu$. We consider 
$L^2(\mu)$ as the reservoir Hilbert space $\mathcal{H}_R$ and introduce the global system-reservoir Hilbert space $\mathcal{H}=\mathcal{H}_S\otimes 
\mathcal{H}_R$. The reservoir space $\mathcal{H}_R$ is equipped with the natural product orthonormal basis $\mathbb{B}$ consisting of all vectors of the form 
$\vert\mathbf{x}\rangle=\vert x_1\rangle\otimes \vert x_2\rangle\otimes \cdots \otimes\vert x_n\rangle\cdots\equiv \vert x_1, x_2,\cdots , x_n\cdots\rangle$, where $\mathbf{x}$ varies over all sequences of elements $x_1,x_2,\cdots$, with only a finite number of nonzero elements from $\mathbb{X}$. 
We single out the state $\vert\Omega_0\rangle=\vert 0\rangle\otimes \vert 0\rangle\otimes \cdots \otimes \vert 0\rangle \cdots$ and call it 
the {\it reservoir vacuum}. Considered as a function on the probability space $(\Omega, \mu)$, the reservoir vacuum state 
$\vert\Omega_0\rangle$ is the constant function, identically equal to unity. 

Denote by $U_{0j}$, the unitary operator in $\mathcal{H}$, determined by its action, 
\begin{equation} 
U_{0j}\, \vert \phi \otimes \mathbf{x}\rangle = \sum_{y\in\mathbb{X}} \left(\, L_{y\, x_j}\, \vert\,\phi\rangle\right)\, \otimes   
\vert x_1,x_2,\cdots , x_{j-1}, y, x_{j+1},\cdots \rangle,\ \ \forall\ \ \vert\phi\rangle\,\in \mathcal{H}_S, \vert\mathbf{x}
\rangle\in\mathbb{B}.  
\end{equation} 
The unitary operator $U_{0j}$ acts essentially on the tensor product of $\mathcal{H}_S$ and the $j^{\rm th}$ copy of $L^2(\nu)$ in 
the reservoir space 
$$\mathcal{H}_R=L^2(\mu)=L^2(\nu)\otimes L^2(\nu)\otimes \cdots $$
where the countable tensor product on the right hand side is with respect to the {\it stabilizing sequence} 
$(\vert 0\rangle, \vert 0\rangle, \cdots )$. Put 
\begin{equation} \label{discreteU}
U_n=U_{0\,n}\, U_{0\,n-1}\, \cdots U_{0\,1},\ \ \ \  n=1,2,\cdots  
\end{equation}     
and $U_0=I$ the identity operator in $\mathcal{H}$. Then $\{U_n\}$ determines a discrete time inhomogeneous Schr{\"o}dinger evolution satisfying 
\begin{equation}
U_n\, \vert \phi \otimes \mathbf{x}\rangle = \sum_{y_1,y_2,\cdots}\,  \left(L_{y_n\,x_n}\, L_{y_{n-1}\,x_{n-1}}\, \cdots  L_{y_1\,x_1}\, \vert \phi\rangle\right) \otimes \vert y_1,y_2\cdots , y_n\rangle \otimes  
\vert x_{n+1},x_{n+2},\cdots \rangle  . 
\end{equation} 
for all states $\vert \phi\rangle$ in $\mathcal{H}_S$ and $\vert \mathbf{x}\rangle\in \mathbb{B}.$
It is clear that  $U_n$ is a unitary operator in $\mathcal{H}$ for every $n$. For any operator $X\in \mathcal{B}(\mathcal{H}_S),$  
 \begin{equation} \label{Tndef}
\langle \chi\vert T^{n}(X)\vert \phi\rangle =\langle \chi\otimes \Omega_0\vert U_n^\dag\, (X\otimes I_R)\, U_n\vert \phi\otimes \Omega_0\rangle,\ \ \forall\,\, \vert \chi\rangle,\vert \phi\rangle\in \mathcal{H}_S.  
\end{equation}    
This admits the following interpretation: The irreversible discrete time dynamics of the system $S$ described by the quantum dynamical semigroup $\{T^n\}$ is obtained by reducing the Heisenberg dynamics of the system observables, induced by the unitary Schr{\"o}dinger dynamics $\{U_n\}$ of the system plus reservoir. This reduction is in the reservoir vacuum state $\vert \Omega_0\rangle$. 

We now look at the evolution of the initial state 
\begin{equation}
\vert\psi_0\rangle=\vert \phi_0\otimes \Omega_0\rangle,\ \  \vert\phi_0\rangle\in \mathcal{H}_S  
\end{equation}  
in $\mathcal{H}$ under $\{U_n\}$ by explicitly expressing
\begin{eqnarray} \label{psin}
\vert\psi_n\rangle &=& U_n\, \vert \phi_0\otimes \Omega_0\rangle \nonumber \\ 
&=& \sum_{y_1,y_2,\cdots , y_n}\ \left( L_{y_n\, 0}\, L_{y_{n-1}\, 0}\, \cdots L_{y_1\, 0}\, \vert\,\phi_0\rangle\right)\otimes 
\vert y_1, y_2,\cdots ,y_n\rangle \otimes \vert 0, 0, \cdots \rangle.      
\end{eqnarray} 
Now, let us consider a measurement on the reservoir, when the global state in $\mathcal{H}$ is  
given by $\vert\psi_n\rangle$ of (\ref{psin}). If we get a classical output $(y_1,y_2,\cdots, y_n)\in \mathbb{X}^n$ as a result of the measurement, the post-measured  state is  
\begin{equation} \label{collapse}
\vert \Psi_n(y_1,y_2,\cdots y_n)\rangle_S=\frac{\, L_{y_n\, 0}\, L_{y_{n-1}\, 0}\, \cdots L_{y_1\, 0}\, \vert\phi_0\rangle }
{\vert\vert L_{y_n\, 0}\, L_{y_{n-1}\, 0}\, \cdots L_{y_1\, 0}\, \phi_0\vert\vert}  
\end{equation}
where $\vert\vert \phi \vert\vert$ denotes norm of the vector $\vert\phi\rangle$ in $\mathcal{H}_S$. Note that whenever the denominator vanishes, it is clear from 
(\ref{collapse}) that the classical output $(y_1,y_2,\cdots, y_n)$ cannot occur. Thus the random collapsed state $\vert \Psi_n(y_1,y_2,\cdots y_n)\rangle_S$ is defined only on the subset 
$$\{(y_1,y_2,\cdots , y_n):\    L_{y_n\, 0}\, L_{y_{n-1}\, 0}\, \cdots L_{y_1\, 0}\, \vert\,\phi_0\rangle\neq 0\}\  \subset \mathbb{X}^n.$$ 
What we have described above is succinctly illustrated  in Fig.~1 in the form of a quantum circuit. 

\begin{figure}
\includegraphics*[width=5in,keepaspectratio]{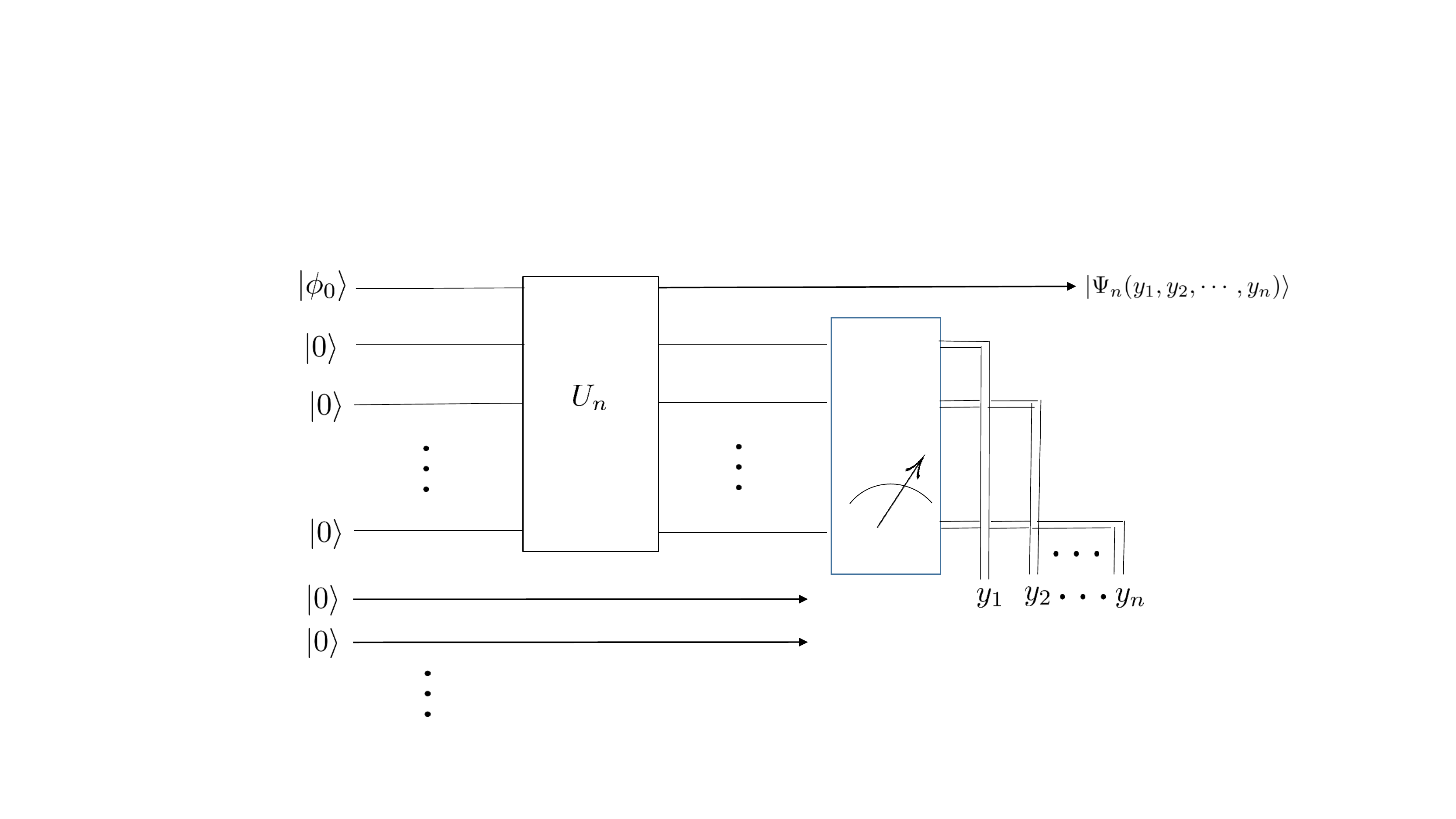}
\caption{Evolution of the initial state $\vert\phi_0\otimes \Omega_0\rangle$  in $\mathcal{H}=\mathcal{H}_S\otimes \mathcal{H}_R$, induced by a unitary operator $U_n$ (see (\ref{discreteU})), followed by a measurement on the reservoir yielding a classical output $(y_1,y_2,\cdots, y_n)$ and a post measured state $\vert \Psi_n(y_1,y_2,\cdots y_n)\rangle_S$ of the system.} 
\end{figure}

Alternatively, allowing a 1-step evolution by $U_{01}$ on the initial state $\vert \phi_0\otimes 0\rangle$ and making a measurement, we  get a classical output $y_1$ and a collapsed state $\vert\Psi_1(y_1)\rangle_S$ of the system $S$ given by 
\begin{equation} 
\label{psiy1}
\vert\Psi_1(y_1)\rangle_S=\frac{L_{y_1\,0}\, \vert\, \phi_0\rangle}{\vert\vert\, L_{y_1\,0}\, \phi_0\vert\vert}. 
\end{equation}
Now, allow this collapsed state to undergo a one-step evolution again, and make a measurement. We get a classical output $y_2$ and a collapsed state $\vert\Psi_2(y_1,y_2)\rangle_S$ given by  
\begin{equation}
\label{psiy1y2} 
\vert\Psi_2(y_1,y_2)\rangle_S= \frac{L_{y_2\,0}\, \vert\Psi_1(y_1)\rangle_S }{\vert\vert\, L_{y_2\,0}\,\Psi_1(y_1)\vert\vert}
=\frac{ L_{y_2\,0}\,L_{y_1\,0}\, \vert\,\phi_0\rangle}{\vert\vert L_{y_2\,0}\,L_{y_1\,0}\, \phi_0\vert\vert}.
\end{equation} 
Repeating this procedure $n$ times, we get a classical output sequence $(y_1,y_2,\cdots , y_n)$ and the collapsed state 
$\vert\Psi_n(y_1, y_2,\cdots ,y_{n})\rangle_S$ of the system, given by the same expression as in (\ref{collapse}). Furthermore, 
 \begin{equation}
\label{psinplus1}
\vert\Psi_{n+1}(y_1, y_2,\cdots ,y_{n+1})\rangle_S=\frac{ \, L_{y_{n+1}\, 0}\, \vert\, \Psi_n(y_1,y_2,\cdots y_n)\rangle_S}
{\vert\vert L_{y_{n+1}\, 0}\, \Psi_n(y_1,y_2,\cdots y_n)\vert\vert}. 
\end{equation} 
Thus the sequence $\{\vert\Psi_n(y_1, y_2,\cdots ,y_{n})\rangle_S, n=1,2,\cdots\}$ of $\mathcal{H}_S$-valued random variables is a Markovian sequence~\cite{Gisin,Carm,Molmer1,Molmer2,Massen2}, adapted to the random trajectory 
$(y_1,y_2,\cdots)$ of the reservoir, occurring as a classical stochastic process in the wake of successive measurements.

Put 
$$\nu_n(y_1,y_2,\ldots, y_n)=\vert\vert L_{y_{n}\, 0} \cdots L_{y_2\,0}\,L_{y_1\,0}\, \phi_0\vert\vert^2$$
and observe that 
$$\sum_{(y_1,y_2,\cdots , y_n)\in \mathbb{X}^n}\, \nu_n(y_1,y_2,\ldots, y_n)=1.$$
Moreover, 
$$\sum_{y_{n+1}\in \mathbb{X}}\, \nu_{n+1}(y_1,y_2,\ldots, y_n,y_{n+1})=\nu_n(y_1,y_2,\ldots, y_n),$$
for $n=1,2,\cdots$. In other words, $\nu_n$ is a probability distribution on $\mathbb{X}^n$, which is also the marginal distribution of $\nu_{n+1}$ on the product of the first $n$ copies of 
$\mathbb{X}$.  Thus, $\{\nu_n\}$ is a consistent family of distributions over $\{\mathbb{X}^n\}$. By Kolmogorov's consistency theorem, there exists a unique probability measure $\nu_\infty$ in the countable product space 
$\Omega=\mathbb{X}^\infty$, whose  marginal on the product of the first $n$ copies of $\mathbb{X}$ is $\nu_n$ for every $n=1,2,\cdots$. The probability measure $\nu_\infty$ describes the statistics of the discrete measurement sequence $(y_1,y_2,\cdots )$.  Putting  
$$Z_n(\mathbf{y})=k^n\, \nu_n(y_1,y_2,\cdots , y_n),\ n=1,2,\cdots, \mathbf{y}=\omega\in\Omega, $$
we obtain the likelihood ratio  martingale sequence $\{Z_n\}$ in the probability space $(\Omega, \mu)$. The sequence $\{Z_n\}$ is a non-negative martingale, with $\mathbb{E}_\mu\, [Z_n]=1$, for all $n$. However, there is no guarantee that $\nu_\infty$ is absolutely continuous with respect to $\mu$. Thus, the martingale $\{Z_n\}$ need not converge to a finite random variable.  A simple computation shows that 
\begin{eqnarray*} \label{finite_m}
\sum_{(y_1,y_2,\cdots , y_n)\in \mathbb{X}^n}\, && \left\vert \Psi_n(y_1,y_2,\cdots , y_n)\right\rangle \left\langle 
\Psi_n(y_1,y_2,\cdots , y_n)   \right\vert\ \nu_n(y_1,y_2,\cdots, y_n)  \nonumber \\    
&=& \int_\Omega\, \left\vert \Psi_n(y_1,y_2,\cdots , y_n)\right\rangle \left\langle\Psi_n(y_1,y_2,\cdots , y_n)   
\right\vert \, \nu_\infty(d\mathbf{y}) \nonumber \\ 
&=& \int_\Omega\, \left\vert \Psi_n(y_1,y_2,\cdots , y_n)\right\rangle \left\langle\Psi_n(y_1,y_2,\cdots , y_n)\right\vert \, Z_n(\mathbf{y})\, \mu(d\mathbf{y}), 
\end{eqnarray*}
 where  $\nu_\infty(d\mathbf{y})$ gets replaced by $Z_n(\mathbf{y})\, \mu(d \mathbf{y})$ for all $n=1,2,\cdots.$ 

This summarizes the way the discrete time irreversible dynamics  is determined by the discrete time state-valued Markov chain 
$\{\vert\Psi_n(\cdot)\rangle\}$ starting from $\vert\phi_0 \otimes \Omega_0\rangle$.  Furthermore, this suggests a natural route for an extension to the continuous time irreversible dynamics described by a quantum dynamical semigroup $\{T_t, t\geq 0\}$ with GKSL generator $\mathcal{L}$. We can replace the discrete Schr{\"o}dinger evolution $\{U_n, n=0,1,2,\cdots\}$ by the HP unitary dilation $\{U(t), t\geq 0\}$ of $\{T_t, t\geq 0\}$ in the tensor product of $\mathcal{H}_S$ with an appropriate Boson Fock space $\mathcal{H}_R$, and transfer it to $\mathcal{H}_S\otimes L^2(\mu)$ with $\mu$ as the Wiener probability measure of a suitable multidimensional Brownian motion $\{\mathbf{B}(t),\, t\geq 0\}$, using the Wiener-It{\^o}-Segal isomorphism. Putting $\vert\psi_t\rangle=U(t)\, \vert \phi_0\otimes \Omega_0\rangle$, with $\vert\phi_0\rangle\in \mathcal{H}_S$, $\vert\Omega_0\rangle$ being the constant function in $L^2(\mu)$, identically equal to unity, and  normalizing $\vert\psi_t\rangle$ in $\mathcal{H}_S$, we shall arrive at a state diffusion process $\{\vert \Psi_t(\mathbf{B})\rangle, t\geq 0\},$ which is a perfect continuous time analogue of the Markov chain $\{\vert \Psi_n(\cdot )\rangle\}$ given by (\ref{psiy1})-(\ref{psinplus1}).

\section{Boson Fock space and quantum stochastic evolutions}
  
We begin with some general observations on the Boson Fock space $\Gamma(\mathfrak{h})$ over a Hilbert space $\mathfrak{h}$ defined by 
\begin{equation}
\label{gammaspace}
\Gamma(\mathfrak{h})=\mathbb{C}\oplus \mathfrak{h} \oplus \mathfrak{h}^{\textcircled{s}^2} \oplus \cdots \oplus \mathfrak{h}^{\textcircled{s}^r}\oplus \cdots 
\end{equation} 
where $\mathbb{C}$ denotes the one dimensional complex Hilbert space and  $\textcircled{s}^r$ indicates $r$-fold symmetric tensor product of copies of $\mathfrak{h}$. To each $u\in \mathfrak{h}$, its associated exponential vector $e(u)$ is defined by 
\begin{equation} \label{eq2}
e (u) = 1 \oplus u \oplus \frac{u^{\otimes^{2}}}{\sqrt{2!}}  \oplus \cdots 
\oplus \frac{u^{\otimes^{r}}}{\sqrt{r!}}  \oplus \cdots. \end{equation}
The linear manifold generated by all such exponential vectors is denoted by $\mathcal{E}$. Any finite set of exponential vectors is linearly independent and $\mathcal{E}$ is dense in $\Gamma(\mathfrak{h})$. This implies that any map from the set of all exponential vectors into 
$\Gamma(\mathfrak{h})$ extends to an operator in $\Gamma(\mathfrak{h})$ with domain $\mathcal{E}$. 
Any isometry on the set of exponential vectors extends to an isometry on $\Gamma(\mathfrak{h})$. The map $u\rightarrow e(u)$ is strongly continuous and for all $u,\,\, v\in \mathfrak{h}$ 
\begin{equation}
\langle e(u)\vert e(v)\rangle = {\rm exp}\langle u\vert v\rangle . 
\end{equation}
Any element of the subspace $\mathfrak{h}^{\textcircled{s}^r}$ in $\Gamma(\mathfrak{h})$ is called an $r$-particle vector. 
The linear manifold $\mathcal{M}$ generated by $\bigcup \mathfrak{h}^{\textcircled{s}^r}$ in  $\Gamma(\mathfrak{h})$ is called the manifold of finite particle vectors. To any $u\in \mathfrak{h}$, there is associated a pair of operators $a(u)$, $a^\dag(u)$, defined on the linear manifold $\mathcal{M}$, which are closable (with their corresponding closures denoted by the same symbols) and  are called the creation-annihilation pairs associated with $u$. Then, $\mathcal{E}$ is contained in the domain of $a(u)$ and $a^\dag(u)$. These operators  are adjoint to each other on $\mathcal{M}$ and $\mathcal{E}$. They enjoy very important properties and the algebra generated by them gives rise to a rich family of observables. 

The map $u\rightarrow a(u)$ is antilinear whereas  $u\rightarrow a^\dag({u})$ is linear.  The operator $a(u)+a^\dag(u)$ closes to a selfadjoint operator and therefore, yields an observable. The linear manifolds $\mathcal{M}$ and $\mathcal{E}$ are in the domain of products of all operators of the form $F_1, F_2, \cdots , F_l$ where each $F_i$ is either $a(u_i)$ or $a^\dag(u_i)$ for each $i=1,2,\cdots l$. On both 
$\mathcal{M}$ and $\mathcal{E}$ the creation and annihilation operators obey the canonical commutation relations: 
\begin{eqnarray} \label{commu}
\ [a(u),\, a(v)]&=& 0,  \nonumber \\ 
\ [a^\dag(u),\, a^\dag(v)] &=& 0, \\ 
\ [a(u),\, a^\dag(v)] &=& \langle u\vert v\rangle. \nonumber
\end{eqnarray} 
Furthermore, 
\begin{equation} \label{acteu}
a(u)\, e(v)=\langle u\vert v\rangle\, e(v), \ \forall\ \ u, v\in \mathfrak{h}. 
\end{equation}
If $\mathfrak{h}_1,\ \mathfrak{h}_2$ are two Hilbert spaces,  the correspondence 
\begin{equation} \label{esumproduct}
e(u_1\oplus u_2)\rightarrow e(u_1)\otimes e(u_2), \ \ \forall\ \, u_i\in\mathfrak{h}_i, i=1,2
\end{equation} 
extends to a Hilbert space isomorphism between $\Gamma(\mathfrak{h}_1\oplus\mathfrak{h}_2)$ and $\Gamma(\mathfrak{h}_1)\otimes\Gamma(\mathfrak{h}_2)$. 

Now we specialize to the case where $\mathfrak{h}=L^2(\mathbb{R}_+,\mathbb{C}^n)=L^2(\mathbb{R}_+)\otimes \mathbb{C}^n$, where $L^2(\mathbb{R}_+)$ is the  Hilbert space of absolutely square integrable functions on the half-interval $\mathbb{R}_+=[0,\infty)$, with respect to the Lebesgue measure  and  $\mathbb{C}^n$  denotes the standard $n$-dimensional complex Hilbert space.  The Hilbert space $L^2(\mathbbm{R}_+)\otimes \mathbb{C}^n$ can be viewed as the space of $\mathbb{C}^n$-valued norm square integrable functions on $\mathbb{R}_{+}$. 
Any element $\mathbf{u}\in L^2(\mathbb{R}_+) \otimes \mathbb{C}^n$ may be expressed as,  
$$\mathbf{u}= u_1\oplus u_2\oplus \cdots \oplus u_n,\ \ \ \ \ \ u_k\in L_2(\mathbb{R}_+),\  k=1,2,\cdots, n.$$
With any $\mathbf{u}\in L^2(\mathbb{R}_+)\otimes \mathbb{C}^n$, we associate the exponential vector $e(\mathbf{u})$ in 
$\Gamma(L^2(\mathbb{R}_+)\otimes \mathbb{C}^n)$. For any $\mathbf{u}$ and $\mathbf{v}$ in $L^2(\mathbb{R}_+)\otimes \mathbb{C}^n$ we have 
\begin{eqnarray}
\langle e(\mathbf{u})\vert e(\mathbf{v})\rangle&=&{\rm exp}\langle \mathbf{u}\vert \mathbf{v}\rangle \nonumber \\ 
                                               &=&{\rm exp}\left[\sum_{k=1}^n\,\int_0^\infty\,  u_k^*\, v_k\, dt\right]. 
																						\end{eqnarray}
The vacuum vector $e(\mathbf{\mathbf{0}}) =   1 \oplus \mathbf{0} \oplus \mathbf{0} \oplus \cdots$ in  $\Gamma(L^2(\mathbb{R}_+)\otimes \mathbb{C}^n)$ is denoted by 
$\Omega_0$.  


We consider a quantum system $S$  in a Hilbert space $\mathcal{H}_S$, coupled to a reservoir $R$     
in a Boson Fock space $\mathcal{H}_R = \Gamma (L^2 (\mathbbm{R}_+) \otimes \mathbb{C}^n)$.  The global Hilbert space $\mathcal{H}=\mathcal{H}_S\otimes\mathcal{H}_R$ is used to describe events, observables and states of the system plus reservoir. The noise processes can be described by observables in the general continuous tensor product Hilbert space of the reservoir for which  the Boson Fock space $\Gamma (L^2 (\mathbbm{R}_+)\otimes \mathbb{C}^n)$ serves as one of the simplest models. The space $\mathbb{C}^n$ corresponds to $n$ degrees of freedom in the selection of noise. 
   
For any $0<t_1<t_2\cdots <t_r<\infty$, we have the following decomposition of $\mathcal{H}=\mathcal{H}_S\otimes \mathcal{H}_R$: 
\begin{eqnarray*}  
\mathcal{H}([0,t))&=&\mathcal{H}_S\otimes \Gamma(L^2([0,t))\otimes \mathbb{C}^n)   \\
 \mathcal{H}([t_{r-1},t_{r})) &=& \Gamma(L^2([t_{r-1},t_{r}))\otimes \mathbb{C}^n) \\
\mathcal{H}([t_r,\infty)) &=& \Gamma(L^2([t_r,\infty))\otimes \mathbb{C}^n) 														
\end{eqnarray*} 
and we denote the restrictions of $\mathbf{u}$ to the time intervals $[0,t]$, $[t_1,t_2)$,  and $[t_r,\infty)$ by 
\begin{eqnarray*}
\left.\mathbf{u}\right\vert_{[0,t)}&=&\mathbf{u}_{t]}, \\ 
\left.\mathbf{u}\right\vert_{[t_{r-1},t_{r})}&=&\mathbf{u}_{[t_{r-1},t_{r})}, \\
\left.\mathbf{u}\right\vert_{[t_r,\infty)}&=&\mathbf{u}_{[t_r}.
\end{eqnarray*} 

From the correspondence given by (\ref{esumproduct}), it follows that, there exists a unique unitary isomorphism $\mathcal{U}: \mathcal{H} \rightarrow \mathcal{H}([0,t_{1})) \otimes 
\mathcal{H}([t_{1}, t_{2})) \otimes \cdots \otimes \mathcal{H}([t_{r-1},t_{r})) \otimes \mathcal{H}([t_{r},\infty))$ satisfying,     
\begin{equation}
\mathcal{U}\, \phi\otimes e(\mathbf{u}) = \phi\otimes e(\mathbf{u}_{t_1]})) \otimes e(\mathbf{u}_{[t_1,t_2)})\otimes \cdots 
\otimes e(\mathbf{u}_{[t_{r-1},t_r)}) \otimes e(\mathbf{u}_{[t_r})   \label{eq4}\\
\end{equation}
 for all $\phi\in \mathcal{H}_S$ and  $e(\mathbf{u})\in \mathcal{H}_R$.

Using the notions of creation and annihilation operators introduced in (\ref{commu}), (\ref{acteu}), we consider the family of linear operators  
$\{A_k(t), t\geq 0\}$ and  $\{A^\dag_k(t), t\geq 0\}$ as follows: 
\begin{eqnarray}
A_k(t)&=&I_S \otimes a \left (1_{[0,t]}  \otimes \vert k\rangle \right ) 
\label{eq6} \\
A^\dagger_k(t) &=& I_S \otimes a^{\dagger} \left (1_{[0,t]}  \otimes \vert k\rangle 
\right )  \label{eq7} 
\end{eqnarray}
where $\{\vert k\rangle = (0, \cdots, 0, 1,0,\cdots,0)\}$ (with $1$ in the $k$-th place), $k=1,2, \cdots, n$, is  a canonical orthonormal basis in $\mathbb{C}^n $;   $1_{[0,t]}$ denotes the indicator function of the interval $[0,t]$ for each $t\in\mathbb{R}_+$ and 
$I_S$  denotes the identity operator in $\mathcal{H}_S$.  
The operators defined in (\ref{eq6}) and (\ref{eq7}) obey the canonical commutation relations (CCRs): 
\begin{eqnarray}
\, [A_k(s),A_l(t)]&=&0=[A^\dag_k(s),A^\dag_l(t)], \label{eq8} \\   
\, [A_k (s),A^\dag_l(t)]&=&\delta_{kl}\, 
(s \wedge t)\, I_S\otimes I_R. \label{eq9}
\end{eqnarray}
Here $s\wedge t$ denotes the minimum of $s$ and $t$. 

The operators $A_k(t),\ A^\dag_k(t)$ are well-defined on the linear manifold generated by elements of the form $\phi 
\otimes e (\mathbf{u}),$ with $\phi \in \mathcal{H}_S$ and $\mathbf{u} \in L^2(\mathbb{R}_+\otimes \mathbb{C}^n)$. In particular, one obtains the following eigen-relation for $A_k(t)$: 
\begin{eqnarray} \label{a_eig}
A_k(t)\,  \vert \phi \otimes e(\mathbf{u})\rangle  
																			&=& \left(\int_{0}^{t} u_k(s)\, ds\, \right)\, \,\vert\phi \otimes e(\mathbf{u})\rangle
\end{eqnarray} 
and consequently, the adjoint relation for $A^\dag_k(t)$ follows: 
\begin{eqnarray} \label{ad_e}
\langle \phi \otimes e(\mathbf{u})  \vert A^\dag_k(t) 
                                           &=& \langle \phi \otimes e(\mathbf{u})  \vert \left(\int_{0}^{t} u^*_k(s)\, ds\, \right).    
\end{eqnarray}
The family of operators $\{A_k(t), t\geq 0\}$, $\{A^\dag_k(t), t\geq 0\}$ are respectively called the annihilation and creation processes. These are the fundamental noise processes of 
quantum stochastic calculus. (For more detailed description of fundamental noise processes in Boson Fock space, including conservation noise process, see Refs.~\cite{HP, KRP}).  


A family $X = \left \{ X(t), 0 \leq t < \infty \right \}$ of operators in  
$\mathcal{H}$ is said to be {\it adapted} if, for each $t,$ there 
exists an operator $X_t$ in $\mathcal{H}([0,t))$ such that $$X(t) = X_t 
\otimes I_{[t}$$ where $I_{[t}$ is the identity operator in $\mathcal{H}([t,\infty)).$  
Further, an adapted process $X$ is said to be {\it simple} with respect to a partition $0 < 
t_1 < t_2 < \cdots < t_r < \cdots $ of $[0, \infty)$ such that $t_r 
\rightarrow \infty$ as $r \rightarrow \infty$, if 
\begin{equation}
X(t)=X({t_j}) \ \  {\rm when}\ t_j\leq t<t_{j+1},\ \ j=0,1,2,\cdots .  \label{eq12}
\end{equation}
Let $\{L(t)\}$ be such a {\it simple adapted process} and $\{M(t)\}$ be any one of the fundamental operator-valued adapted processes 
$\{A_k(t)\}$, $\{A_k^\dag(t)\},\ k=1,2,\cdots , n$. Then,  
the stochastic integral of   $\{L(t)\}$, with respect to  $\{M(t)\}$  is defined by  
\begin{eqnarray}  \label{eq13}
X(t)&=&\int_{0}^t\,  L(s)\, d\, M(s)  \nonumber \\
 &=&\sum_{t_j}\, L_{t_j}\, \left( M (t_{j+1} \wedge t)
- M (t_{j} \wedge t)\right),   \label{eq11} \\   
&&  \ \ \ \ \ \ \ \ \ \ \ t_j\leq t<t_{j+1}, \ \ \ j=0,1,2,\cdots . \nonumber   
\end{eqnarray}
It may be noted that the operators $L_{t_j}$ and  $M (t_{j+1} \wedge t)- M (t_{j} \wedge t)$ commute with each other i.e., $ L(s)\, d\, M(s)$ can be written as $d\, M(s)\, L(s).$ 

As shown in Ref.~\cite{HP}, the notion of such integrals can be extended by a completion procedure to a wide class of adapted processes, which are not necessarily simple. Such an integration is a linear operation in the space of adapted processes.  For details see Sec.~4 of Ref.~\cite{HP}. 
 
We consider adapted processes of the form  
\begin{eqnarray} \label{xteq}
X(t)=X(0)+ \int_0^t\, \sum_{k=1}^n\, \left(E_{k}(s)\, dA^\dag_k(s)\, + F_{k}(s)\, dA_k(s)+ G_{k}(s)\, ds\right)  \label{eq14}
\end{eqnarray}    
where $X(0)=X_0\otimes I_R$,  $X_0$ is an operator in the system Hilbert space $\mathcal{H}_S$ and $I_R$ denotes the identity operator in 
$\mathcal{H}_R$; the integrands $E_k(t), F_k(t), G_k(t)$ are adapted processes. 
We write (\ref{xteq})  in the differential form as, 
\begin{eqnarray}
dX(t)=\sum_{k=1}^n\, \left( E_{k}(t) dA^\dag_k(t)\, + F_{k}(t)\, dA_k(t)+\,  G_{k}(t)\, dt\right),  \label{eq15}
\end{eqnarray}    
with initial value $X_0\otimes I_R$. 

The central result of quantum stochastic calculus is the following quantum It{\^o}  multiplication table~\cite{HP, KRP}, summarized as follows: 

\begin{equation}    \label{eq16}
\begin{tabular}{c||c|c|c}\
 &\hskip 0.1in $dA^\dag_k \hskip 0.1in$   &\hskip 0.1in $dA_k$ \hskip 0.1in & \hskip 0.1in $dt$ \hskip 0.1in \\ 
\hline 
\hline
\hskip 0.1in $dA^\dag_l$ \hskip 0.1in & 0 & 0 & 0 \\ 
\hline 
$dA_l$ & $\delta_{kl}\ dt$ & 0 & 0 \\ 
\hline 
$dt$ & 0 & 0 & 0 \\ 
\hline

\end{tabular}
\end{equation}

The product of two stochastic integrals is again a stochastic integral, the  differentials of which satisfy the modified Leibnitz relation, 
\begin{equation}
d(X\,Y)= (dX)\, Y+ X\, (dY)+\, (dX)\, (dY).  \label{eq17}  
\end{equation} 
Quantum It\^{o} multiplication table (\ref{eq16})  is employed in (\ref{eq17}) to express the differential $d(X\,Y)$ of the product of  adapted processes $X$, $Y$ in terms of the fundamental operator-valued differentials $dA^\dag_k,\, dA_k$ and 
$dt$.  This provides a simple and natural extension of It\^{o}  calculus based on Brownian motion~\cite{McKean} to  its quantum counterpart in the Boson Fock space. 

One of the most successful applications of  HP quantum stochastic calculus is the realization of unitary dilations of quantum dynamical 
semigroups through Schr{\"o}dinger evolutions of open systems. Such a Schr{\"o}dinger evolution can be expressed through a {\it unitary operator-valued process} obeying a quantum stochastic differential equation of the form,
\begin{equation}
dU(t) =\left(\sum_{k=1}^n\, \left(L^{(1)}_k\, dA^\dag_k(t) +
L^{(2)}_k\, dA_k(t) \right) + L^{(3)}\, dt \right)\! U(t),\ \ \ U(0)=I \label{eq18}  
\end{equation} 
in  $\mathcal{H}=\mathcal{H}_S\otimes \mathcal{H}_R$, 
where  $L^{(\alpha)}_k,\ \alpha=1,2,3$ are bounded operators in $\mathcal{H}_S$. It is shown~\cite{HP} that a unique unitary solution for (\ref{eq18}) exists if  
\begin{eqnarray}
L^{(1)}_k&=&L_k, L^{(2)}_k=-L^\dag_k\nonumber \\
L^{(3)}_k&=&-i\, H - \frac{1}{2}\, \sum_{k=1}^n\, L_k^\dag\, L_k \label{eq19}  
\end{eqnarray}
and $H$ is a self-adjoint operator.  Taking the conditions (\ref{eq19}) into account, (\ref{eq18}) can be expressed as~\cite{HP, KRP}  
\begin{equation}  \label{eq21}
dU(t) =\left[\sum_{k=1}^n\, \left(L_k\, dA^\dag_k(t) -
L^\dag_k\, dA_k(t)\right) -\left(i\, H+ \frac{1}{2}\, \sum_{k=1}^n\, L_k^\dag\, L_k \right) \, dt \right]\! U(t),\ \ \ U(0)=I, \end{equation}
which is referred to as the HP equation. In terms of the set of operators $\mathbf{L}=(L_1,L_2,\cdots, L_n)$ and $H$, we denote the unitary process $\{U(t), t\geq 0\}$ satisfying (\ref{eq21}) by  $U\! (\mathbf{L}, H)$. 
In the special case of $L_k=0$ for all $k=1,2,\cdots, n$, one obtains the familiar Schr{\"o}dinger unitary dynamics 
\begin{equation} \label{eq20}
dU(t)=-i\, H\, U(t),  
\end{equation}
with $H$ being the Hamiltonian of the quantum system.    It is of interest to note that there do exist examples with unique unitary solutions, when the coefficients $L_k$ and $H$ in (\ref{eq21}) are unbounded~\cite{Fan, Fan&Wills, Fan13, KRPG}.

We may now use the unitary process $\{U(t), t\geq 0\}$ to describe  noisy Heisenberg dynamics. To this end, consider any bounded operator $X$ in the system 
Hilbert space $\mathcal{H}_S$ (i.e., $X\in \mathcal{B}(\mathcal{H}_S)$),  and a unitary process $U\!(\mathbf{L}, H)$. Define 
 a homomorphism $j_t:  \ \mathcal{B}(\mathcal{H}_S)\longrightarrow 
\mathcal{B}(\mathcal{H}_S\otimes\mathcal{H}_R)$ by 
\begin{equation}   \label{eq22}
 j_t (X) = U(t)^{\dagger} (X \otimes I_R)  U(t), \quad t \ge 0.   
\end{equation}
Using the relation (\ref{eq17}) and employing the quantum It\^{o} multiplication table given by (\ref{eq16}),  one obtains 
\begin{equation}   \label{eq23}
dj_t(X)= \sum_{k=1}^{n}\, \left\{\, j_t\left([X, L_k]\right)\, dA^\dag_k(t)  - j_t\left([X,L_k^\dag]\right)\, dA_k(t) \right\} 
         +j_t\left(\mathcal{L}(X)\right)\, dt,   
\end{equation}
where   the map $\mathcal{L}$  from $\mathcal{B}(\mathcal{H}_S)$ to itself is given by 
\begin{eqnarray}
\mathcal{L}(X)&=& i\left[ H,\,  X\right]- \frac{1}{2} \sum_{k=1}^{n}\left(L^\dag_k\, L_k\, X + X\, L^\dag_k\, L_k -2\, L^\dag_k\, X\, L_k\right)  \label{eq24}
\end{eqnarray} 
Equation (\ref{eq23}) describes noisy evolution of system observables $X$.   
If  $L_{k}=0 \ \ \ \forall\ \,k,$ then (\ref{eq23}) 
reduces to the well-known Heisenberg equation of motion for the observable $X$:
$$\frac{d j_t (X)}{dt} = j_t (i [H,X]).$$

 For any operator $F$ in $\mathcal{H}$ we define the 
{\it vacuum conditional expectation value} as the unique operator $\mathbb{E}_{\Omega_0} (F)$ in $\mathcal{H}_S$ determined by,
\begin{equation} \label{vacexp}
\langle \phi\vert \mathbb{E}_{\Omega_0} (F) \vert \chi\rangle= \langle \phi\otimes \Omega_0 |F| \chi \otimes \Omega_0\rangle,\ \ \forall\,\, \ \phi,\chi \in \mathcal{H}_S.
\end{equation}
Now, we write the vacuum conditional expectation value of $j_t (X)$ as   
\begin{eqnarray}  \label{condexp}
\mathbb{E}_{\Omega_0} \left(j_t(X)\right)&=&\mathbb{E}_{\Omega_0}\, \left(U(t)^{\dagger} (X \otimes I_R)  U(t)\right)\nonumber \\ 
&=&T_t(X)
\end{eqnarray} 
Thus one obtains 
\begin{equation}  \label{eq26}
\frac{d T_t (X)}{dt} = T_t (\mathcal{L}(X))=\mathcal{L}\left(T_t(X)\right) 
\end{equation} 
for the time evolution of the quantum dynamical semigroup of completely positive unital maps 
\begin{equation}
T_t={\rm exp}( t\, \mathcal{L}),\ \  t\geq 0
\end{equation}  
on $\mathcal{B}(\mathcal{H}_S)$ generated by  $\mathcal{L}$ of (\ref{eq24}).  This coincides with the well-known form obtained by 
Gorini, Kossakowski, Sudarshan~\cite{Gorini} and Lindblad~\cite{Lindblad}.  

For the initial state $\rho_0\otimes \vert \Omega_0\rangle\langle\Omega_0\vert$ of the system plus reservoir, we express,
\begin{eqnarray}   \label{eq27}
{\rm Tr}\left(\rho_0\otimes \vert \Omega_0\rangle\langle\Omega_0\vert\, j_t(X)\right)&=& {\rm Tr}\left(\rho_0\, T_t(X)\right) \nonumber \\ 
&=& {\rm Tr}\left(\rho_t\, X\right),
\end{eqnarray} 
where $\rho_t={\rm Tr}_R\left( U(t)\, \rho_0\otimes \vert \Omega_0\rangle\langle\Omega_0\vert\, U(t)^\dag\right)$ denotes the reduced density operator of the quantum system. Using (\ref{condexp})-(\ref{eq27}) we get the Gorini-Kossakowski-Sudarshan-Lindblad (GKSL) master equation for $\rho_t$ :
\begin{equation}
 \frac{d \rho_t}{dt} = -i [ H,\,\rho_t]- \frac{1}{2} \sum_{k=1}^{n}\left(L^\dag_k\, L_k\, \rho_t + \rho_t\, L^\dag_k\, L_k -2\, L_k\, \rho_t\, L^\dag_k\right).  \label{eq28}
\end{equation}  
 In the next section we discuss invariance properties of the GKSL generator $\mathcal{L}$.   

\section{Symmetries of the GKSL generator} 

Let $\{R_i(t)=I_S\otimes F_i(t), t\geq 0\},\  i=1,2$  be unitary adapted processes in $\mathcal{H}$ such that  $\{F_i(t), t\geq 0\},\, i=1,2$, act only on the reservoir space $\mathcal{H}_R$. Let 
\begin{eqnarray}
F_2(t)\vert\, \Omega_0\rangle=F_2(t)^\dag\,\vert \Omega_0\rangle=\vert\Omega_0\rangle\, ,\ \ t\geq 0. 
\end{eqnarray}
Consider the process 
\begin{equation} \label{vt}
\{V(t)=R_1(t)\, U(t)\, R_2(t),\  t\geq 0\}
\end{equation} 
where  $U(t)$ satisfies the HP equation (\ref{eq21}). Define
 a homomorphism $j'_t:  \ \mathcal{B}(\mathcal{H}_S)\longrightarrow 
\mathcal{B}(\mathcal{H}_S\otimes\mathcal{H}_R)$ by 
\begin{eqnarray}
 j'_t (X) = V^{\dagger}(t) (X \otimes I_R)  V(t), \quad t \ge 0.   
\end{eqnarray}
 Then, the vacuum conditional expectation value (see (\ref{vacexp}) and (\ref{condexp})) of $j'_{t}(X)$ is given by, 
\begin{eqnarray}
\mathbb{E}_{\Omega_0} \left(j'_t(X)\right)
&=& \mathbb{E}_{\Omega_0}\, \left(R^{\dagger}_2(t)\, U^\dag(t)\, R^\dag_1(t) (X \otimes I_R)  R_1(t)\, U(t)\, R_2(t)\right)\nonumber \\
&=& \mathbb{E}_{\Omega_0}\,\left( U^\dag(t)\,  (X \otimes I_R)  U(t)\, \right)=\mathbb{E}_{\Omega_0} \left(j_t(X)\right)=T_t(X)
=e^{t\,\mathcal{L}}(X).  
\end{eqnarray} 
for all $t\geq 0$ and $X$ in $\mathcal{B}(\mathcal{H}_S)$. Thus, conjugation by the unitary adapted processes $\{U(t)\}$ and $\{V(t)\}$ yield the reduced dynamics of the quantum system with  the same GKSL generator $\mathcal{L}$. 
In the following, we discuss two important examples of $\{V(t), t\geq 0\}$, which specialize to {\em the  translation  and rotation invariance} of the GKSL generator $\mathcal{L}$. 

\subsection{Example 1} 

In analogy with exponential vectors of (\ref{eq2}) we now introduce {\it exponential operators}  in $\mathcal{H}_R$ as follows:  For any $\mathbf{f}\in L^2(\mathbb{R}_+)\otimes \mathbb{C}^n$, we write, on the set of exponential vectors, 
\begin{equation}
W(\mathbf{f}) e(\mathbf{u}) = e^{-\frac{1}{2} ||\mathbf{f}||^{2} - \langle \mathbf{f}|\mathbf{u} \rangle} e (\mathbf{f}+\mathbf{u}) \quad \forall\, \ \mathbf{u}\in\mathcal{K},      \label{eq29}
\end{equation} 
where $\vert\vert\mathbf{f}\vert\vert^{2}=\int_0^{\infty} \vert \mathbf{f}\vert^2\, dt,$ and $\vert \mathbf{f}\vert^2= \sum_{k=1}^{n}\,\vert  f_k\vert^2$.  
The exponential operator $W(\mathbf{f})$ preserves the scalar product between exponential vectors and therefore extends to a unique unitary operator in $\mathcal{H}_R$, which we denote by the same symbol $W(\mathbf{f})$. 

A normalized vector $\alpha(\mathbf{f})\in\mathcal{H}_R$  given by    
\begin{eqnarray} \label{coherent}
\alpha(\mathbf{f})&=&W(\mathbf{f})\, e(\mathbf{0}) =  e^{-\frac{1}{2} ||\mathbf{f}||^{2}}\, e (\mathbf{f}), \ \ 
\mathbf{f}\in L^2(\mathbb{R_+})\otimes \mathbb{C}^n,    
\end{eqnarray} 
is called a {\it coherent state} associated with $\mathbf{f}$.  

The operators $W(\mathbf{f}),\ W(\mathbf{g})$ obey the multiplication relation, 
\begin{equation} \label{eq30}
 W(\mathbf{f}) W(\mathbf{g}) = e^{-i \,\,{\rm Im} \langle \mathbf{f}|\mathbf{g} \rangle}\, W(\mathbf{f}+\mathbf{g}),\ \ \forall\, \ \mathbf{f},\mathbf{g}\in \mathcal{K}. 
\end{equation}
These are the well known Weyl canonical commutation relations (CCRs) of which the CCRs of creation and annihilation operators (
\ref{commu}) are the infinitesimal versions. We call  $W(\mathbf{f})$ the Weyl displacement operator associated with $\mathbf{f}.$ 

Now, for any map $\mathbf{f}:\mathbb{R}_+\rightarrow \mathbb{C}^n$ satisfying the local square integrability condition 
$$\int_0^{t}\,   \vert \mathbf{f}(s)\vert^2 \, ds < \infty, \ \forall\, \ t>0 $$
we introduce the unitary {\it Weyl displacement operator process} $\{W(\mathbf{f})(t) , t\geq 0\}$ by the relation  
\begin{equation} \label{eq31}
W(\mathbf{f})(t)\, e(\mathbf{u})= W(1_{[0,t]} \mathbf{f})\, e(\mathbf{u}_{t]}) \otimes e(\mathbf{u}_{[t}).
\end{equation} 
Then $\{R_{\mathbf{f}}(t)=I_S\otimes W(\mathbf{f})(t), t\geq 0\}$ is a unitary adapted  process in $\mathcal{H}_S\otimes \mathcal{H}_R$, which obeys the quantum stochastic differential equation 
\begin{equation} \label{eq32}
dR_{\mathbf{f}}(t)=\left\{\sum_{k=1}^n\, \left(f_k\, dA^\dag_k(t) - f_k^* dA_k(t) \right) -\frac{1}{2}\,\sum_{k=1}^n\, \vert f_k\vert^2\, dt \right\}\! R_{\mathbf{f}}(t),\ \ t\geq 0    
\end{equation} 
with  initial condition $R_{\mathbf{f}}(0)=I_S\otimes I_{R}$. 

Choose $R_1(t)=R_{\mathbf{f}}(t)$, $R_2=I_S\otimes I_R$ in 
(\ref{vt}).  Then, $\{V(t)=R_{\mathbf{f}}(t)\,U(t), t\geq 0\}$,  is a unitary adapted process satisfying     
\begin{eqnarray}  \label{eq33}
dV(t)&=& \left[ dR_{\mathbf{f}}(t)\right] U(t)+ R_{\mathbf{f}}(t) \left[ dU(t)\right]+\left[ dR_{\mathbf{f}}(t) \right]\,  
\left[dU(t)\right] \nonumber \\ 
 &=& \left\{\sum_{k=1}^n\, \left((L_k+f_k)\, dA^\dag_k(t) - (L^\dag_k+f^*_k)\, dA_k(t) \right) \right.\nonumber \\ 
  && \hskip 0.5in \left. - \left(iH+\frac{1}{2}\, 
\sum_{k=1}^n\,(L_k^\dag L_k+ \vert f_{k}\vert^2 + 2\, f_k^*L_k\right) \right\}\! V(t) 
 \end{eqnarray}    
with initial condition $V(0)=I_S\otimes I_R$. The process $\{V(t), t\geq 0\}$ is, indeed, given by 
$$\{V(t), t\geq 0\}=U\left(\mathbf{L}',\, H'\right),$$
where  $\mathbf{L}'=\mathbf{L}+\mathbf{f}$ and $H'=H+\frac{1}{2i}\sum_{k=1}^n\, \left(f_k^*L_k-f_k\, L_k^\dag\right).$    

Clearly, the homomorphism $j_{t,\mathbf{f}}:  \ \mathcal{B}(\mathcal{H}_S)\longrightarrow 
\mathcal{B}(\mathcal{H}_S\otimes\mathcal{H}_R)$ defined by  
$$j_{t,\mathbf{f}} (X) = V(t)^{\dagger} (X \otimes I_R)  V(t)$$
satisfies the relation 
$$j_{t,\mathbf{f}} (X)\equiv U(t)^{\dagger} (X \otimes I_R)  U(t)=j_t(X)$$  
and hence,  the generator $\mathcal{L}$,  defined by (\ref{eq24}) with operators $(\mathbf{L},\, H)$, remains invariant, when $\mathbf{L}$, $H$ are replaced by $\mathbf{L}'=\mathbf{L}+\mathbf{f}$ and $H'=H+\frac{1}{2i}\, 
\displaystyle\sum_{k=0}^{n}(f_k^*L_k-f_k\, L_k^\dag)$ respectively.        

\noindent {\em Remark}: When $\mathbf{f}(\cdot)$ is a constant vector $\pmb{\ell}$ for all $t\geq 0$, it follows that $\mathbf{L}'=\mathbf{L}+\pmb{\ell}$ and $H'=H+\frac{1}{2i}\, \displaystyle\sum_{k=0}^{n}(\ell_k^*L_k-\ell_k\, L_k^\dag)$, thereby exhibiting the {\em translation invariance} property of the GKSL generator $\mathcal{L}$. 
\medskip 

\subsection{Example 2} 

Let $t\rightarrow \mathbf{F}(t)$ be an $n\times n$  unitary matrix-valued Borel map on $\mathbb{R}_+$. Define the second quantization unitary operator process $\{\Gamma(\mathbf{F})(t), t\geq 0\}$, acting only on $\mathcal{H}_R$, by the relation 
\begin{equation} \label{euzeta}
\Gamma(\mathbf{F})(t)\, e(\mathbf{u})=\Gamma(\mathbf{F})(t)\,e(u\otimes \pmb{\zeta})= 
e\left(u_{t]}\otimes \mathbf{F}(t)\, \pmb{\zeta}\right)\otimes e(u_{[t}\otimes \pmb{\zeta}).  
\end{equation}
 where we  use the identification $L^2(\mathbb{R}_+,\mathbb{C}^n)=L^2(\mathbb{R}_+)\otimes\mathbb{C}^n$ and choose 
$\mathbf{u}=u\otimes \pmb{\zeta}$ with $u \in  L^2(\mathbf{R}_+)$ and  $\pmb{\zeta}\in\mathbb{C}^n$. Then, 
\begin{equation} \label{gammaf}
\Gamma(\mathbf{F})(t)\, \Omega_0= \Gamma^\dag(\mathbf{F})(t)\, \Omega_0=\Omega_0.   
\end{equation}
Define 
\begin{equation}
R(t)=I_S\otimes \Gamma(\mathbf{F})(t),\ \ \  t\geq 0. 
\end{equation}
and choose $R_1(t)=R(t)$, $R_2(t)=R^\dag(t)$ in (\ref{vt}). Then, 
$$V(t)=R(t)\,U(t)\, R^\dag(t),\  t\geq 0.$$  Define the homomorphism 
$j_{t,\mathbf{F}}:  \ \mathcal{B}(\mathcal{H}_S)\longrightarrow 
\mathcal{B}(\mathcal{H}_S\otimes\mathcal{H}_R)$ by 
\begin{eqnarray}
j_{t,\mathbf{F}} (X) &=& V^{\dagger}(t) (X \otimes I_R)  V(t) \nonumber \\
     &=& I_S\otimes \Gamma(\mathbf{F})(t)\,   U^{\dagger}(t) (X \otimes I_R)  U(t)\, I_S\otimes \Gamma^\dag(\mathbf{F})(t),\ \forall \ \  t\geq 0. 
\end{eqnarray}  
Then, it follows immediately from  (\ref{gammaf}) that, 
\begin{eqnarray}
\mathbb{E}_{\Omega_0} \left(j_{t,\mathbf{F}}(X)\right)&=&\mathbb{E}_{\Omega_0}\, \left(U^{\dagger}(t) (X \otimes I_R)  U(t)\right)\nonumber \\ 
&=& \mathbb{E}_{\Omega_0} \left(j_t(X)\right)=e^{t\,\mathcal{L}}(X).
\end{eqnarray}
In other words, both   $\{U(t)\}$ and $\{V(t)=R(t)\,U(t)\, R^\dag(t)\}$ yield the irreversible dynamics of the states and observables of the quantum system with the same GKSL generator $\mathcal{L}$.

\noindent {\em Remark}: Consider a special case of the second quantization unitary process $\{\Gamma(\mathbf{F}(t)\}$, where $\mathbf{F}(t)$ is a constant $n\times n$ unitary matrix defined by, $\mathbf{F}(t)=((u_{ij})),\ i,j=1,2,\cdots , n$ for all $t\geq 0$. Then, $\{U(t), t\geq 0\}=U(\mathbf{L},H)$ and 
$\{V(t), t\geq 0\}=U(\mathbf{L}',H')$, where $L'_i=\sum_{j=1}^n\, u_{ij}\, L_j, \ \ H'=H$.  The GKSL generator $\mathcal{L}$ remains invariant, when  the operator parameters $(\mathbf{L}, H)$ are replaced by $(\mathbf{L'}, H')$, thereby exhibiting the  {\em rotation invariance} property of $\mathcal{L}$.

\section{Wiener-It\^{o}-Segal isomorphism}

We shall now describe the HP quantum stochastic calculus in the Hilbert space $L^2(\mu)$, where $\mu$ is the classical Wiener  probability measure of the $n$-dimensional standard Brownian motion process $\{\mathbf{B}(t), t\geq 0\}$. To this end, we denote 
 $\left\{\mathbf{B}(t)^T=\left(B_1(t), B_2(t),\cdots , B_n(t)\right)^T\right\}$ where $B_k(t),$ $1\leq k\leq n$ are $n$ independent one dimensional standard Brownian motion processes, `$T$' denoting transpose. We introduce the exponential random variables 
\begin{eqnarray} \label{eq36}
\widetilde{e}(\mathbf{u})(\mathbf{B})= {\rm exp}\,\left( \int_0^\infty \mathbf{u}(s)^T\, d\mathbf{B}(s)- 
\frac{1}{2}\,\int_0^\infty \mathbf{u}(s)^T\mathbf{u}(s)\, ds\right),\ \ \mathbf{u}\in L^2(\mathbb{R}_+)\otimes\mathbb{C}^n, 
\end{eqnarray}         
where we view $L^2(\mathbb{R}_+)\otimes \mathbb{C}^n$ also as the direct sum of $n$ copies of $L^2(\mathbb{R}_+).$  
Now, consider the correspondence 
$$\Theta :\ e(\mathbf{u}) \rightarrow \widetilde{e}(\mathbf{u}),$$
where $e(\mathbf{u})$ is the exponential vector  defined in Section~III (see (\ref{eq2})). 
The map $\Theta$ is scalar product preserving and so, it extends uniquely to a Hilbert space isomorphism from the Boson Fock space 
$\Gamma(L^2(\mathbb{R}_+)\otimes\mathbb{C}^n)$ to  $L^2(\mu)$. This is called the Wiener-It\^{o}-Segal isomorphism~\cite{Wiener, Ito, Segal}.  

For any vector $\phi$ in $\mathcal{H}_R=\Gamma(L^2(\mathbb{R}_+)\otimes\mathbb{C}^n)$ or in 
$\mathcal{H}=\mathcal{H}_S\otimes\mathcal{H}_R$, we write 
\begin{eqnarray} \label{eq37}
\widetilde{\phi}=\left\{ \begin{array}{l} 
                   \Theta\, \phi, \ {\rm if}\ \phi\in \mathcal{H}_R, \\ 
									I_{S}\otimes \Theta\, \phi , \ {\rm if}\  \phi\in \mathcal{H}.
                       \end{array} \right.
\end{eqnarray}    
Then $\phi\rightarrow \widetilde{\phi}$ is a Hilbert space isomorphism from $\mathcal{H}_R \rightarrow L^2(\mu)$  as well as  
$\mathcal{H}\rightarrow \mathcal{H_S}\otimes L^2(\mu)$.  We shall identify $\mathcal{H_S}\otimes L^2(\mu)$ with  
 the space $L^2(\mu,\mathcal{H}_S)$ of $\mathcal{H}_S$-valued norm square integrable functions on the space of Brownian paths.  A typical element of 
$L^2(\mu,\mathcal{H}_S)$ is a functional $\widetilde{\phi}(\mathbf{B})$ and the scalar product of two vectors $\widetilde{\phi}_1$, $\widetilde{\phi}_2$ in 
$L^2(\mu,\mathcal{H}_S)$ is given by, 
\begin{eqnarray} \label{eq38} 
\langle \widetilde{\phi}_1\vert \widetilde{\phi}_2\rangle &=& \mathbb{E}_{\mathbf{B}}[\langle\,  \widetilde{\phi}_1\vert \widetilde{\phi}_2\rangle_S]
=\int\, \langle \widetilde{\phi}_1(\mathbf{B})\vert \widetilde{\phi}_2(\mathbf{B})\rangle_S\ 
\mu(d\mathbf{B})
\end{eqnarray}  
where $\langle \cdot \vert \cdot \rangle_S$ denotes scalar product in the system Hilbert space $\mathcal{H}_S$ and 
$\mathbb{E}_{\mathbf{B}}[\cdot ]$ denotes  expectation value with respect to  $\mu$.  For any operator $X$ in $\mathcal{H}_R$ or $\mathcal{H}$, we write 
$$\widetilde{X}=\Theta\, X\, \Theta^{-1}.$$ 
Denote by $\mu_{[t_1,t_2]}$, $\mu_{[t_1,\infty)},$  the probability measure of the Brownian motion
$$\{\mathbf{B}(t+t_1)-\mathbf{B}(t_1),\ \ 0\leq t\leq 
t_2-t_1\}.$$  
It may be noted that the factorizability property 
\begin{equation} \label{eq39}
L^2(\mu)=L_2(\mu_{[0,t_1]})\otimes L_2(\mu_{[t_1,t_2]})\otimes \cdots \otimes L_2(\mu_{[t_{r-1},t_r]})\otimes L_2(\mu_{[t_r,\infty)}), 
\end{equation} 
holds for all $0< t_1 <t_2 <\cdots <t_{r-1}< t_r < \infty$.  In other words, the  isomorphism $\Theta$ between $\mathcal{H}_R$ and $L^2(\mu)$ preserves the continuous tensor product structure. With the restriction of $\mathbf{u}\in L^2(\mathbb{R}_+)\otimes\mathbb{C}^n$ to the time interval  $[t_1, t_2]$,  $0\leq t_1 <t_2 <\infty$, in $\mathbb{R}_+$,    
the exponential random variables in $L^2(\mu_{[t_1,t_2]})$ are expressed by   
\begin{eqnarray} \label{e_time}
\widetilde{e}(\mathbf{u}_{[t_1,t_2]})(\mathbf{B})=
{\rm exp}\,\left( \int_{t_1}^{t_{2}} \mathbf{u}(s)^T\, d\mathbf{B}(s)- 
\frac{1}{2}\,\int_{t_1}^{t_{2}} \mathbf{u}(s)^T\mathbf{u}(s)\, ds\right).   
\end{eqnarray}   
The Wiener-It\^{o}-Segal isomorphism maps the vacuum vector $\Omega_0=e(\mathbf{0})$ of the Boson Fock space  
to the constant function in $L^2(\mu)$, identically equal to unity.  Furthermore, we have the following proposition, 
which identifies the sum of creation and annihilation processes in $\Gamma(L^2(\mu)\otimes \mathbb{C}^n)$  with multiplication by components of the $n$-dimensional Brownian motion    
 in $L^{2}(\mu)$ under the isomorphism $\Theta$.     

\noindent{\bf Proposition:}  Let 
$$Q_k(t)=A_k(t)+A^\dag_k(t),\ \  0\leq t<\infty$$ 
in  $\Gamma(L^2(\mathbb{R}_+)\otimes \mathbb{C}^n)$. Then, $\Theta\, Q_k(t)\, \Theta^{-1}$ is {\em multiplication by Brownian motion} random variable $B_k(t)$ in  $L^{2}(\mu)$ i.e.,  
\begin{equation}  \label{eq40}
 [\, \widetilde{Q}_k(t)\,\, \widetilde{\phi}\,]\, (\mathbf{B})= B_k(t)\, \widetilde{\phi}\,(\mathbf{B})   
\end{equation}
for all $\widetilde{\phi}\in L^2(\mu, \mathcal{H}_S)$ under the Wiener-It{\^o}-Segal isomorphism.  
  
\noindent{\bf Proof:} 	Using (\ref{a_eig}), (\ref{ad_e}), we obtain 
\begin{eqnarray} \label{Qproc}
\langle e(\mathbf{u})\vert Q_k(t)\vert e(\mathbf{v})\rangle &=& e^{\langle \mathbf{u}\vert \mathbf{v}\rangle}\,\int_{0}^{t}\, (u^*_k+v_k)(s)\, ds, 
 \end{eqnarray}
 which yields, 
\begin{eqnarray} \label{Qdif}
\frac{d}{dt}\,\langle e(\mathbf{u})\vert Q_k(t)\vert e(\mathbf{v})\rangle &=& 
e^{\langle \mathbf{u}\vert \mathbf{v}\rangle}\ \, (u^*_k+v_k)(t)
 \end{eqnarray}
in $\Gamma(L^2(\mathbb{R}_+\otimes \mathbb{C}^n).$

On the other hand,   
\begin{eqnarray} \label{BkIso} 
\mathbb{E}_{\mathbf{B}}\,\left[B_k(t)\,\{\widetilde{e}(\mathbf{u})^{*}\}\, \{\widetilde{e}(\mathbf{v})\}\right]
&=& e^{\langle \mathbf{u}\vert \mathbf{v}\rangle}\, \mathbb{E}_{\mathbf{B}}\, \left[B_k(t)\, {\rm exp}\left\{\beta_{u_k^*+v_k}(t)\right\}\right] 
\end{eqnarray}  
where  $\beta_{u_k^*+v_k}(t)$ satisfies   
\begin{equation} \label{exp_sde}
d\, \beta_{u_k^*+v_k}(t)= (u^*_k+v_k)(t)\, dB_k(t) - \frac{1}{2} \, (u^*_k+v_k)^2(t)\, dt. 
\end{equation}
Simple application of classical It{\^o} calculus~\cite{McKean} leads to
\begin{eqnarray} \label{last}
\frac{d}{dt}\, \left(\mathbb{E}_{\mathbf{B}}\,\left[B_k(t)\,\{\widetilde{e}(\mathbf{u})^{*}\}\, \{\widetilde{e}(\mathbf{v})\}\right]\right)&=& e^{\langle \mathbf{u}\vert \mathbf{v}\rangle}\, (u^*_k+v_k)(t), 
\end{eqnarray}
thus establishing the proposition.  \hskip 2.5in $\square$
 
We shall now explain how the Weyl displacement process $\{W(\mathbf{f})(t), t\geq 0\}$, discussed in Section IV, looks like in $L^2(\mu)$. 
Under the $\Theta$ isomorphism $W(\mathbf{f})(t)$ satisfies the relation   
\begin{eqnarray}   \label{Weyl}
\widetilde{W}(\mathbf{f})(t)\widetilde{e}(\mathbf{u})(\mathbf{B})&=&\widetilde{e}(\mathbf{u}+1_{[0,t]}\, \mathbf{f})(\mathbf{B}) \, \, 
\times {\rm exp}\left[-\frac{1}{2}\, \int_0^t\, \vert \mathbf{f}(s)\vert^2\, ds -\int_0^t\, \mathbf{f}^\dag\mathbf{u}(s)\, ds\right] \nonumber \\
&=&\widetilde{e}(\mathbf{u}_{[t})(\mathbf{B})\,\, e^{\gamma_{\mathbf{u}}(t,\mathbf{B})}. 
\end{eqnarray}     
where $\gamma_{\mathbf{u}}(t,\mathbf{B})$ is a non-anticipating Brownian functional, obeying  
\begin{eqnarray} \label{beta}  
d\gamma_{\mathbf{u}}&=&  (\mathbf{f}+\mathbf{u})^T\, d\mathbf{B}-\frac{1}{2}\,\left[ \mathbf{f}^\dag\mathbf{f}\, +
(\mathbf{f}+\mathbf{u})^T(\mathbf{f}+\mathbf{u})+2\,\mathbf{f}^\dag\mathbf{u} \right]\, dt.
\end{eqnarray}   
This suggests the possibility of introducing a {\it randomized} Weyl displacement operator $\widetilde{\mathbb{W}}(\mathbf{f})(t)$ by replacing 
$\mathbf{f}(t)$ by a {\it non-anticipating} Brownian functional $\mathbf{f}(t,\mathbf{B})$ in 
(\ref{Weyl}) and  (\ref{beta}). To this end, we consider 
the class  
$$\mathcal{F}_2 = \{\mathbf{f}: \mathbf{f}=\mathbf{f}(t,\mathbf{B}), \int_{0}^{t}\, \vert \mathbf{f}(s,\mathbf{B})\vert^2\, ds 
<\infty \ \forall\ t\geq 0\}$$  
of non-anticipating  $\mathbb{C}^n$-valued   Brownian functionals.  
For any $\mathbf{f}\in \mathcal{F}_2$, we define
\begin{eqnarray} \label{eq42}
\widetilde{\mathbb{W}}(\mathbf{f})(t)\widetilde{e}(\mathbf{u})(\mathbf{B})&=&
\widetilde{e}(\mathbf{u}_{[t})(\mathbf{B})\, e^{\hat{\gamma}_\mathbf{u}(t)} 
\end{eqnarray}  
where the differential of $\hat{\gamma}_\mathbf{u}(t)$ obeys  (\ref{beta}), with $\mathbf{f}\in \mathcal{F}_2$.  We shall now prove that the randomized Weyl displacement operators $\widetilde{\mathbb{W}}(\mathbf{f})(t)$ are unitary.  

\noindent{\bf Theorem:} For any $\mathbf{f}$ in $\mathcal{F}_2$, the family  
$\{\widetilde{\mathbb{W}}(\mathbf{f})(t), t\geq 0\}$ is  a unitary operator-valued adapted process.

\noindent{\bf Proof:}  Substituting (\ref{eq42}) we get,  
\begin{eqnarray}  \label{unitary_W1}
\langle \widetilde{\mathbb{W}}(\mathbf{f})(t)\, \widetilde{e}(\mathbf{u})\vert \widetilde{\mathbb{W}}
(\mathbf{f})(t)\, \widetilde{e}(\mathbf{v})\rangle &=& \mathbb{E}_{\mathbf{B}}\, \left[ \left\{{\rm exp}\left({\hat{\gamma}_\mathbf{u}^*(t)}
+{\hat{\gamma}_\mathbf{v}(t)}\right) \right\}
\langle \widetilde{e}(\mathbf{u}_{[t})\,\vert  \widetilde{e}(\mathbf{v}_{[t})\,\rangle\, \, \right] \nonumber \\
&=&  \mathbb{E}_{\mathbf{B}}\, \left[   \left\{{\rm exp}\left({\hat{\gamma}_\mathbf{u}^*(t)}
+{\hat{\gamma}_\mathbf{v}(t)}\right) \right\} {\rm exp}\left(\int_t^\infty\, \mathbf{u}^\dag\mathbf{v}\, dt\right)\right]
\end{eqnarray}
where
 $\hat{\gamma}_u^*,\ \hat{\gamma}_v$ obey (\ref{beta}), but with  $\mathbf{f}$ in $\mathcal{F}_2.$
On simplification using standard classical It{\^o} calculus~\cite{McKean} we obtain  
\begin{eqnarray}
d\langle \widetilde{\mathbb{W}}(\mathbf{f})(t)\, \widetilde{e}(\mathbf{u})\vert \widetilde{\mathbb{W}}
(\mathbf{f})(t)
\, \widetilde{e}(\mathbf{v})\rangle&=& 0.
\end{eqnarray}
thus establishing that the random Weyl process is unitary in $L^2(\mu)$.   \hskip 1.3in   $\square$

In a similar vein  consider an $n\times n$ unitary matrix-valued nonanticipating Brownian functional $\{\mathbf{F}(t,\mathbf{B}), t\geq 0\}$ and introduce the randomized second quantization process $\{\widetilde{\Gamma}(\mathbf{F})(t), t\geq 0\}$ by the following relation: 
\begin{eqnarray}
\widetilde{\Gamma}(\mathbf{F})(t)\, \widetilde{e}(\mathbf{u})&=& {\rm exp}\left(\int_0^t\, \mathbf{F}(s,\mathbf{B})\mathbf{u}(s)\cdot d\mathbf{B}(s) - \frac{1}{2}\, \int_0^t\, \mathbf{F}(s,\mathbf{B})\mathbf{u}(s) \cdot \mathbf{F}(s,\mathbf{B})\mathbf{u}(s)\, ds\,\right)  
\otimes \widetilde{e}(\mathbf{u}_{[t}), \nonumber \\   && \hskip 0.7in \ t\geq 0, \mathbf{u}\in L^2(\mathbb{R}_+)\otimes \mathbb{C}^n.  
\end{eqnarray}   
Then, a simple algebra, using the It{\^o} calculus, shows that $\{\widetilde{\Gamma}(\mathbf{F})(t), t\geq 0\}$ is scalar product preserving on the set of exponential vectors in $L^2(\mu)$ and hence, determine a {\em randomized second quantization} unitary process, which can be transferred to an adapted unitary process in the Boson Fock space through the Wiener-It{\^o}-Segal isomorphism. 

We shall present some applications of randomized Weyl displacement and randomized second quantization processes in a separate article.

\noindent{\em Remark:} For every  $t\geq 0$ one obtains a {\em Randomized coherent state} 
$\alpha(\mathbf{f})(t)=\mathbb{W}(\mathbf{f})(t)\, e(\mathbf{0})$  
where $\mathbf{f}\in\mathcal{F}_2$. 
Then, under $\Theta$ isomorphism, we obtain               
\begin{eqnarray} \label{cohprocess}
\widetilde{\alpha}(\mathbf{f})(t,\mathbf{B})= \Theta\, \alpha(\mathbf{f})(t)  
 = {\rm exp}\left\{\int_0^t \mathbf{f}(s)^T d\mathbf{B}(s)-\frac{1}{2}\,\int_0^t\left[\mathbf{f}(s)^\dag \mathbf{f}(s) +
\mathbf{f}(s)^T \mathbf{f}(s)\right]\, ds\right\},
\end{eqnarray}  
which satisfies,   
\begin{equation}
 d\,\widetilde{\alpha}(\mathbf{f})(t)=[\mathbf{f}(t)^T d\mathbf{B}(t)-\frac{1}{2}\,\mathbf{f}(t)^\dag \mathbf{f}(t)\,  dt]\, 
\widetilde{\alpha}(\mathbf{f})(t),\ \ \widetilde{\alpha}(\mathbf{f})(0)=1. 
\end{equation}
It is interesting to note that $\widetilde{\alpha}(\mathbf{f})(t),\ t\geq 0$ is a {\em randomized coherent 
state-valued non-anticipating Brownian functional}, for each $\mathbf{f}\in \mathcal{F}_2$. The classical stochastic process $\{\widetilde{\alpha}(\mathbf{f})(t),\ t\geq 0\}$  will be used, in the next section, to derive  the quantum state diffusion equation  from the HP equation.

 
\section{Gisin-Percival state diffusion equation from  HP unitary evolution} 

Consider the HP unitary process 
\begin{equation} \label{HPGisinU}
U(\mathbf{L}\oplus i\mathbf{L}, H)=\{U(t), t\geq 0\}
\end{equation}
in $\mathcal{H}_S\otimes \Gamma(L^2(\mathbb{R}_+)\otimes (\mathbb{C}^{n}\oplus \mathbb{C}^{n})))$, where   
$\mathbf{L}=(L_1,L_2,\cdots, L_n)$. Here  $L_k,\, k=1,2,\cdots, n$ and $H$ are  bounded operators 
in  $\mathcal{H}_S$, with $H$ being selfadjoint.  We denote the annihilation and creation processes 
 in the Boson Fock space $\Gamma(L^2(\mathbb{R}_+)\otimes (\mathbb{C}^{n}\oplus \mathbb{C}^{n}))$ by 
$\{A_{\alpha,k} \  
A^\dag_{\alpha,k}, \alpha=1,2,\ k=1,2,\cdots, n\}$. The unitary process $\{U(t)\}$ of (\ref{HPGisinU}) obeys the HP  equation,
\begin{eqnarray} \label{eq62}
dU(t)&=& \left\{ \sum_{k=1}^{n}\, \left(L_k\, dA_{1,k}^\dag(t) - L^\dag_k\, dA_{1,k}(t)+ i\,L_k\, dA_{2,k}^\dag(t) + iL^\dag_k\, 
dA_{2,k}(t)\right)\right.\nonumber \\ 
&& \hskip 1.5in  \left. -\left(i\, H+\sum_{k=1}^{n}L^\dag_k L_k\right) dt\ \right\}\, U(t),
\end{eqnarray}  
 with initial condition $U(0)=I_S\otimes I_R.$  
Let $\phi_0$ be a unit vector in $\mathcal{H}_S$ and let $\Omega_0$ be the vacuum vector in  $\Gamma(L^2(\mathbb{R}_+)\otimes (\mathbb{C}^{n}\oplus \mathbb{C}^{n})$. 
Denote 
\begin{eqnarray} \label{upsi}
U(t)\, \vert\phi_0\otimes \Omega_0\rangle&=&\vert\psi_t\rangle. 
\end{eqnarray}
Since $U(t)$ acts in $\mathcal{H}(t])$, whereas the creation, annihilation differentials  
$dA^\dag_{\alpha,k}(t), dA_{\alpha, k}(t), \alpha=1,2; k=1,2,\cdots, n$ operate in $\mathcal{H}([t, t+dt]),$  
it follows that  $U(t)$  commutes with $dA^\dag_{\alpha,k}(t), dA_{\alpha, k}(t)$. Furthermore, 
$dA_{\alpha, k}\,\vert\Omega_0\rangle=0$. Hence, using (\ref{eq62}) and (\ref{upsi}), we obtain
\begin{eqnarray} \label{psieq1} 
d\,\vert\psi_t\rangle&=&\sum_{k=1}^{n}\, \left[L_k\,  dA_{1,k}^\dag(t) \vert\psi_t\rangle 
+ i\,L_k\,  
dA_{2,k}^\dag(t) \vert\psi_t\rangle\right] -\left[i\, H+\sum_{k=1}^{n}L^\dag_k L_k\right]\vert\psi_t\rangle \, dt \,  \nonumber \\
\end{eqnarray}  
with initial value $\vert\psi_0\rangle=\vert\phi_0\otimes \Omega_0\rangle$. Once again, since $dA_{\alpha, k}\,\vert\psi_t\rangle=0$, we can write (\ref{psieq1}) in terms of   
$\{Q_{\alpha,k}(t)=A_{\alpha,k}(t)+A^\dag_{\alpha,k}(t)\}$ as follows:    
\begin{eqnarray} \label{psieq2} 
d\,\vert\psi_t\rangle&=&\sum_{k=1}^{n}\, \left[L_k\,  dQ_{1,k}(t)\, \vert\psi_t\rangle 
+ i\,L_k\,  dQ_{2,k}(t)\, \vert\psi_t\rangle\right]  -[i\, H+\sum_{k=1}^{n}L^\dag_k L_k ] \, \vert\psi_t\rangle\, dt.
\end{eqnarray}  
Under the Wiener-It{\^o}-Segal isomorphism  
$Q_{\alpha\,k}(t)\rightarrow \Theta\, Q_{\alpha\,k}(t)\Theta^{-1}=\widetilde{Q}_{\alpha,k}(t)$ is multiplication by  $B_{\alpha,k}(t),\ \forall \ \ \,\, t\in\mathbb{R}_+$ in  $L^2(\mu)$ (see proposition of Section V). We replace  the $2n$ dimensional Brownian path $\{B_{\alpha,k}, \alpha=1,2;\, k=1,2,\cdots, n\}$  by the corresponding $n$-dimensional complex Brownian path $\mathbf{B}=\{B_{1,k}+i\,B_{2,k},k=1,2,\cdots , n\}.$ The map defined by  $t\rightarrow \vert\widetilde{\psi}_t(\mathbf{B})\rangle= \Theta U(t)\, \vert\phi_0\otimes \Omega_0\rangle$ is a non-anticipating $\mathcal{H}_S$-valued Brownian functional in $L^2(\mu, \mathcal{H}_S)$, with $\mu$ denoting the Wiener probability measure of $n$-dimensional complex Brownian motion $\mathbf{B}.$
Hereafter, all our discussions will take place in $L^2(\mu, \mathcal{H}_S)$ and we shall omit the symbol  
`$_{\,\,\widetilde\,\, }$' over vectors as well as operators.  

The functional $\vert\psi_t(\mathbf{B})\rangle$ obeys a {\em linear} classical stochastic differential equation 
\begin{equation} \label{psi}
d\, \vert\psi_t\rangle=  \sum_{k=1}^n L_k \vert\psi_t\rangle\, d\,B_k(t)-( i\, H + \sum_{k=1}^n\, L^\dag_k\, L_k)\, \vert\psi_t\rangle\, dt.    
\end{equation}      
 The system density operator 
\begin{equation}
\rho_t=\mathbbm{E}_\mathbf{B}\,[\,\vert\psi_t\rangle\langle \psi_t\vert\,]=\int\, 
\vert\psi_t(\mathbf{B})\rangle\langle\psi_t(\mathbf{B})\vert\, \mu(d\mathbf{B}),
\end{equation}
obtained after {\em coarse graining} over the Brownian paths,  obeys the  GKSL master equation 
\begin{eqnarray}\label{GKSL2}
\frac{d \rho_t}{dt} = -i [H,\, \rho_t]- \sum_{k=1}^{n}\left(L^\dag_k\, L_k\, \rho_t + \rho\, L^\dag_k\, L_k - 2\, L_k\, \rho_t\, L^\dag_k\right). 
\end{eqnarray}
The solution $\vert\psi_t\rangle$ of linear stochastic Schr{\"o}dinger equation (\ref{psi}) does not, in general, have unit norm in $\mathcal{H}_S$. Hence, it  does not result in  a {\em quantum state diffusion}. Using the classical It{\^o} multiplication rule~\cite{McKean}
$$dB_k\,dB_l=0, \ dB_k dB^*_l=2\,\delta_{kl}\, dt,\ \ (dt)^2=0$$ 
for the product of differentials, we obtain 

\begin{eqnarray} \label{norm} 
d\, \langle \psi_t\vert \psi_t\rangle_S &=&  (\langle \psi_t\vert) 
\left(\, d\, \vert\psi_t\rangle\right) + \left(\,d\, \langle \psi_t\vert\right)\,  (\vert \psi_t\rangle)
+ \left(\,d\, \langle \psi_t\vert\right)\, \left(\,d\, \vert \psi_t\rangle\right)  \nonumber \\ 
&=& \sum_{k=1}^{n}\left\{\langle \psi_t\vert L_k\vert\psi_t\rangle_S\,\, dB_k(t) + 
\langle \psi_t\vert L_k^\dag\,\vert\psi_t\rangle_S\ dB_k^*(t)\right\} \nonumber \\ 
&=& 2\, {\rm Re}\, \left[\sum_{k=1}^n\,\langle \psi_t\vert L_k\,\vert\psi_t\rangle_S\, dB_k(t)\,\right].
\end{eqnarray}   
 
Define 
\begin{eqnarray}
\ell_{k,\,\psi_t}=\left\{\begin{array}{l} \frac{\langle \psi_t\vert\, L_k\,\vert\psi_t\rangle_S}
{\langle \psi_t\vert \psi_t\rangle_S},\  {\rm if}\ \langle \psi_t\vert \psi_t\rangle_S\neq 0 \\
 \langle \psi_0\vert\, L_k\,\vert\psi_0\rangle_S,\ \ {\rm otherwise,}
\end{array}
\right. 
\end{eqnarray}
for $k=1,2,\cdots , n$.  Then, $\vert  \ell_{k,\,\psi_t} \vert \leq \vert\vert L_k\,\vert\vert$ and hence, $\ell_{k,\, \psi_t}$
is a non-anticipating Brownian functional in  $\mathcal{F}_2$ . Thus,    
\begin{eqnarray} 
 d\, \langle \psi_t\vert \psi_t\rangle_S&=& 
2\, {\rm Re}\, \left[\sum_{k=1}^{n}\, \ell_{k\, \psi_t} 
dB_k(t)\right]\, \langle \psi_t\vert \psi_t\rangle_S.  
\end{eqnarray}   
This implies 
\begin{eqnarray} \label{psinorm2}
 \langle \psi_t\vert \psi_t\rangle_S &=& 
{\rm exp}\,\left\{\int_{0}^{t}\, 2\, {\rm Re}\, \left[\sum_{k=1}^{n}\,\ell_{k,\psi_s}\, 
dB_{k}(s)\right]\, -\, 2\,\int_{0}^{t}\, \sum_{k=1}^n\, \vert\,\ell_{k,\,\psi_s}\vert^2\, ds\right\}
 \nonumber \\ 
&=& {\rm exp}\,\left\{\int_{0}^{t}\, 2\, {\rm Re}\, 
\left[\sum_{k=1}^n\, \frac{\langle \psi_s\vert L_k\vert\psi_s\rangle_S}{\langle \psi_s\vert \psi_s\rangle_S}\, 
dB_k(s)\right] -\, 2\,\int_{0}^{t}\, \sum_{k=1}^n\, \left|\frac{\langle \psi_s\vert 
L_k\vert \psi_s\rangle_S}{\langle \psi_s\vert \psi_s\rangle_S}\right|^2\, ds\right\}
\nonumber \\ 
&=&  {\rm exp}\,\left\{\int_{0}^{t}\,  2\,{\rm Re}\, \left[\sum_{k=1}^n\,\langle L_k\rangle_{\psi_s}\, 
dB_k(s)\right]   
-\, 2\,\int_{0}^{t}\, \sum_{k=1}^n\,\left\vert \langle L_{k,\psi_s} \right\vert^2\, ds 
\right\}\, 
\end{eqnarray}
where we have denoted $\langle L_k\rangle_{\psi_s}=\frac{\langle \psi_s\vert L_{k}\,\vert\psi_s\rangle_S}
{\langle \psi_s\vert \psi_s\rangle_S}$ in the last line of (\ref{psinorm2}). 

Consider the following exponential classical stochastic process (see (\ref{cohprocess})) in the probability space $(\Omega, \mu)$:     
\begin{equation}\label{rancohproc} 
\{\alpha(\mathbf{f}\oplus i\mathbf{f})(t,\mathbf{B})=\mathbb{W}(\mathbf{f}\oplus i\mathbf{f})(t)\, e(\mathbf{0})(\mathbf{B}),\ 
\mathbf{f}\in\mathcal{F}_2,\, t\geq 0\}.
\end{equation}

Such a process obeys the following classical stochastic differential equation   
			\begin{equation} \label{dcohff}
			d\alpha(\mathbf{f}\oplus i\mathbf{f})=   
			\left\{\sum_{k=1}^{n}\,\left[ f_k\, dB_k-\vert\, f_k\,\vert^2 dt\right]\right\}\, \alpha(\mathbf{f}\oplus i\mathbf{f}).
			\end{equation}
From (\ref{rancohproc}) and (\ref{dcohff}) it may be noted that $\alpha(\mathbf{f}\oplus i\mathbf{f})(t,\mathbf{B})$ is a  non-anticipating Brownian functional.   
We consider a related process    
$\left\{\Phi_t(\mathbf{f})=
{\rm exp}\left[\int_0^t\, 2\,\vert\,\mathbf{f}(s)\,\vert^2 ds\right]\, \alpha(\mathbf{f}\oplus i \mathbf{f})(t,\mathbf{B}), t\geq 0\right\}$ which satisfies 
\begin{equation} \label{dphiff}
			d\Phi_t(\mathbf{f})=   
			\left\{\sum_{k=1}^n\, \left[\, f_k\, dB_k+\vert\,f_k\vert^2 dt\right]\right\}\, \Phi_t(\mathbf{f}). 
			\end{equation}
 \noindent {\bf Theorem}: Let $\vert\psi_t\rangle,\ t\geq 0$ be given by the linear stochastic differential equation (\ref{psi}) and let  
\begin{equation} \label{PsiState}
\vert \Psi_t\rangle=\Phi_t(-\langle\mathbf{L}\rangle_{\psi_t})\, \vert\psi_t\rangle. 
\end{equation}
Then, $\{\vert \Psi_t\rangle, t\geq 0\}$ is an $\mathcal{H}_S$ state-valued diffusion process, which obeys the diffusion equation  
\begin{eqnarray} \label{dPsi2} 
d\,\vert\Psi_t\rangle&=& \sum_{k=1}^n\, (L_k-\langle L_k\rangle_{\Psi_t})\, \vert\Psi_t\rangle\, dB_k(t) - 
\left[i\, H + \sum_{k=1}^n\, \left(L_k^\dag\, L_k-\left|\langle L_k\rangle_{\Psi_t}\right|^2\right)\right] \vert\Psi_t\rangle\, dt.
\end{eqnarray}

\noindent{\bf Proof:} From (\ref{psi}), (\ref{psinorm2}), (\ref{dphiff}) and (\ref{PsiState}) it can be recognized that   
\begin{equation} \label{Psiph}
\vert\Psi_t\rangle=
 \vert \psi_t\rangle\, {\rm exp}\left\{-\int_{0}^t\, \sum_{k=1}^n\, \left[\langle  
L_k\rangle_{\psi_s} d B_k(s)
+ \left|\langle  L_k\rangle_{\psi_s}\right|^2\, ds\right]\, \right\}.   
\end{equation} 
is a normalized vector in $\mathcal{H}_S$. Thus, it immediately follows that      
\begin{equation} \label{defch}
\langle L_k\rangle_{\psi_t}
=\langle L_k\rangle_{\Psi_t},\ \forall\ k=1,2,\cdots n.
\end{equation}
Substitituting (\ref{defch}) in (\ref{Psiph}) 
and applying It{\^o}'s differentiation rules~\cite{McKean} to simplify the differential of (\ref{Psiph}), we obtain the quantum state diffusion equation (\ref{dPsi2}). \hskip 1.5 in $\square$ 

\noindent {\bf Corollary: (Gisin-Percival state diffusion)} The state diffusion equation (\ref{dPsi2}) is equivalent to the  Gisin-Percival quantum state diffusion equation 
\begin{eqnarray} \label{dPsiPG}
 d\vert\Psi_t\rangle&=&\sum_{k=1}^n\, \left(L_k-\langle\, L_k\rangle_{\Psi_t}\right)\, \vert\Psi_t\rangle\, \  dB'_k(t) \nonumber \\ 
&&\ \ \ \ - 
\left(i\, H + \sum_{k=1}^{n}\, \left[ L_k^\dag\, L_k+
\left|\langle L_k\, \rangle_{\Psi_t}\right|^2 -2\, L_k\, \langle\, L_k\rangle^*_{\Psi_t}\right]\right)\, \vert \Psi_t\rangle\, dt, 
\end{eqnarray}
where  $\mathbf{B}'=(B'_1, B'_2, \cdots , B'_n)$ is a process defined by     
\begin{equation}\label{GirB}
dB_k'(t)=dB_k(t) - 2\, \langle L_k\rangle_{\Psi_t}^* dt,\ \ B'_k(0)=0,\ \forall\ k=1,2,\cdots n.   
\end{equation}  
Then, $\mathbf{B}'$ is also a standard Brownian motion in the probability space $(\Omega, \mu_G)$, where (Girsanov's theorem~\cite{Girsanov}), 
\begin{eqnarray}\label{mu2mug}
\mu_G(d\mathbf{B})&=&{\rm exp}\left\{\sum_{k=1}^{n}\, \int_0^t\, 2\,{\rm Re}\left[\langle L_k\, \rangle_{\Psi_s}\, dB_k(s)\right] -2\, 
\int_0^t \vert\,\langle L_k\, \rangle_{\Psi_s}\vert^2\, ds \right\}\, \mu(d\mathbf{B}) \nonumber \\
&=& \langle \psi_t\vert \psi_t\rangle_S\, \mu(d\mathbf{B})  
\end{eqnarray}
for every finite time interval $[0,t].$

\noindent{\bf Proof:} This is immediate from the Girsanov's theorem~\cite{Girsanov}. (Second line of (\ref{mu2mug}) follows from 
(\ref{psinorm2}) and (\ref{defch})). \hskip 1.5in $\square$
  
\noindent{\it Remark:} Note that the factor $\langle \psi_t\vert \psi_t\rangle_S$ in (\ref{mu2mug}) 
appearing under Girsanov measure transformation from $\mu((d\mathbf{B}))$ to  $\mu_G(d\mathbf{B})$,  is a continuous time analogue of the discrete time martingale sequence $\{Z_n\}$ of Sec. II. 
      
It is interesting to note that $\{\vert\Psi_t\rangle, t\geq 0\}$ of (\ref{Psiph}) is, indeed, an explicit solution of Gisin-Percival state diffusion equation (\ref{dPsiPG}). The system density operator $\rho_t, t\geq 0$ is obtained by coarse-graining over the $\mathcal{H}_S$ state-valued trajectories $\{\vert\Psi_t(\mathbf{B}')\rangle$ i.e.,   
\begin{eqnarray} 
\rho_t&=&\mathbb{E}_{\mathbf{B}'}[\vert\Psi_t\rangle\langle\Psi_t\vert] \nonumber \\ 
      		&=&\, \int \vert\psi_t(\mathbf{B})\rangle\langle\psi_t(\mathbf{B})\vert\,  
			\mu(d\mathbf{B}) \nonumber \\ 
									&=& \mathbb{E}_{\mathbf{B}}[\vert\psi_t\rangle\langle\psi_t\vert]
\end{eqnarray}
Evidently, the GKSL master equation (\ref{GKSL2}) obeyed by  $\rho_t$ follows as a consequence of this unraveling. In fact, one may realize different forms for state diffusion processes associated with a GKSL master equation (\ref{GKSL2}), when the operator parameters $(\mathbf{L}, H)$ are replaced by $(\mathbf{L}', H')$ corresponding to symmetry transformations discussed in Sec.~IV. In other words, a single noisy unitary Schr{\"o}dinger evolution driven by a quantum stochastic differential equation (\ref{eq62}) of the HP type results in various forms of Gisin-Percival state diffusion processes associated with a GKSL generator $\mathcal{L}$ of the one parameter quantum dynamical semigroup  $\{T_t, t\geq 0\}$ describing the irreversible dynamics of the  quantum system.

\section{Summary}

We have derived a non-linear stochastic Schr{\"o}dinger equation (\ref{dPsi2}) describing classical diffusive trajectories, with values on the unit sphere of the system Hilbert space $\mathcal{H}_S$, driven by a complex vector-valued standard Brownian motion $\{\mathbf{B}(t), t\geq 0\}$, starting from the quantum stochastic differential equation (\ref{eq62}) of the HP type. This is facilitated by making use of  Wiener-It{\^o}-Segal isomorphism between the reservoir Boson Fock space and 
the Hilbert space $L^2(\mu)$ of norm square integrable functions, with respect to the Wiener probability measure $\mu$ of 
a vector-valued Brownian motion. Consequently, the Gisin-Percival state diffusion equation (\ref{dPsiPG}) is obtained by changing the Brownian motion with an appropriate Girsanov measure transformation. A striking feature of our approach is that it leads to an explicit solution (\ref{Psiph}) of the Gisin-Percival equation in terms of the HP unitary process and a randomized Weyl displacement process. It follows that the system density matrix $\rho_t$, obtained by averaging over the Gisin-Percival diffusive trajectories, obeys a GKSL master equation (\ref{GKSL2}), describing the irreversible dynamics of  states and  observables of the quantum system. Furthermore, it follows that, starting from a single noisy Schr{\"o}dinger unitary evolution (\ref{eq62}) of the HP type, different forms of Gisin-Percival state diffusion processes could be realized, based on the symmetries of the GKSL generator $\mathcal{L}$ of the one parameter quantum dynamical semigroup $\{T_t, t\geq 0\}$.    

\section*{Acknowledgements}
We thank Professor John Gough for pointing out a misprint in Eq.~(\ref{dPsiPG}) of the first version of our manuscript and  Professor Harish Parthasarathy for bringing to our notice an error in the factor multiplying $\mu(d\mathbf{B})$ in Eq.~(\ref{mu2mug}). ARU acknowledges the local hospitality of Indian Statistical Institute, Delhi, where this work was carried out during her sabbatical leave period. She also thanks the University Grants Commission (UGC), India for support under a Major Research Project (Grant No. MRPMAJOR-PHYS-2013-29318).

\end{document}